\documentclass[allclo]{FBSart}
\usepackage{amsfonts}
\usepackage{amssymb}
\usepackage{graphicx} 
\usepackage{clpsref}

\newcommand{\sfrac}[2]{\mbox{\footnotesize $\displaystyle \frac{#1}{#2}$}}
 
\newcommand{\lsim}{\mathrel{\rlap{\lower4pt\hbox{\hskip0pt$\sim$}} 
\raise1pt\hbox{$<$}}}           %less than or approx. symbol 
\newcommand{\gsim}{\mathrel{\rlap{\lower4pt\hbox{\hskip0pt$\sim$}} 
\raise1pt\hbox{$>$}}}           %greater than or approx. symbol 

\title{On Nucleon Electromagnetic Form Factors} 

\author{R.\ Alkofer,\instnr{1} A.\ H\"oll,\instnr{2} 
M.\ Kloker,\instnr{1} A.\ Krassnigg\instnr{2} 
and C.\,D.\ Roberts\instnr{2,3}} 

\instlist{
Institut f\"ur Theoretische Physik, Universit\"at T\"ubingen,
Auf der Morgenstelle 14 \\\mbox{\hspace*{0.9em}D-72076 T\"ubingen, Germany}
\and
Physics Division, Argonne National Laboratory, Argonne, IL
60439-4843, USA
\and
Fachbereich Physik, Universit\"at Rostock, D-18051 Rostock, 
Germany}

\runningauthor{R.\, Alkofer, et al.}
\runningtitle{On Nucleon Electromagnetic Form Factors}
\begin{document}

\maketitle 
\begin{abstract}
A Poincar\'e covariant Faddeev equation, which describes baryons as composites of confined-quarks and -nonpointlike-diquarks, is solved to obtain masses and Faddeev amplitudes for the nucleon and $\Delta$.  The amplitudes are a component of a nucleon-photon vertex that automatically fulfills the Ward-Takahashi identity for on-shell nucleons.  These elements are sufficient for the calculation of a quark core contribution to the nucleons' electromagnetic form factors.  An accurate description of the static properties is not possible with the core alone but the error is uniformly reduced by the incorporation of meson-loop contributions.  Such contributions to form factors are noticeable for $Q^2 \lesssim 2\,$GeV$^2$ but vanish with increasing momentum transfer.  Hence, larger $Q^2$ experiments probe the quark core.  The calculated behaviour of $G_E^p(Q^2)/G_M^p(Q^2)$ on $Q^2\in [2,6]\,$GeV$^2$ agrees with that inferred from polarisation transfer data.  Moreover, $\sqrt{Q^2} F_2(Q^2)/F_1(Q^2) \approx\,$constant on this domain.  These outcomes result from correlations in the proton's amplitude.  
%On the domain of $Q^2$ explored there is no compelling evidence that $F_2(Q^2)/F_1(Q^2) \propto [\ln Q^2/\Lambda^2]^2/Q^2$.
%
\vspace*{1ex}

%\received{{\bf\large PRELIMINARY} \dotfill\ \today \dotfill\ {\bf\large DRAFT}}
\received{\today}
\end{abstract}
%%%%%%%%%%%%%%%%%%%%%%%%%%%%%%%%%%%%%%%%%%%%%%%%%%%%%%%%%%%%%%%%%%%%%%%%%%%
%%%%%%%%%%%%%%%%%%%%%%%%%%%%%%%%%%%%%%%%%%%%%%%%%%%%%%%%%%%%%%%%%%%%%%%%%%%

\section{Introduction}
Modern, high-luminosity experimental facilities that employ large momentum transfer reactions are providing remarkable and intriguing new information on nucleon structure \cite{gao,leeburkert}.  For an example one need only look so far as the discrepancy between the ratio of electromagnetic proton form factors extracted via Rosenbluth separation and that inferred from polarisation transfer \cite{jones,roygayou,gayou,arrington,qattan}.  This discrepancy is marked for $Q^2\gsim 2\,$GeV$^2$ and grows with increasing $Q^2$.  At such values of momentum transfer, $Q^2 > M^2$, where $M$ is the nucleon's mass, a veracious understanding of these and other contemporary data require a Poincar\'e covariant description of the nucleon.  

This is apparent in applications of relativistic quantum mechanics; e.g., Refs.~\cite{millerfrank,boffi,fuda,stan,bruno}.  A different tack follows the formulation of a Poincar\'e covariant Faddeev equation \cite{regfe,hugofe}.  Its foundation is understood through the observation that the same interaction which describes colour-singlet mesons also generates quark-quark (diquark) correlations in the colour-$\bar 3$ (antitriplet) channel \cite{regdq}.  While diquarks do not survive as asymptotic states \cite{mandarvertex}; i.e., they do not appear in the strong interaction spectrum, the attraction between quarks in this channel grounds a picture of baryons in which two quarks are always correlated as a colour-$\bar 3$ diquark pseudoparticle, and binding is effected by the iterated exchange of roles between the bystander and diquark-participant quarks.  

Reference~\cite{cjbfe} reported a rudimentary study of this Faddeev equation and subsequently more sophisticated analyses have appeared; e.g., Refs.\,\cite{bentzfe,oettelfe,hechtfe,birsefe}.  It has become apparent that the dominant correlations for ground state octet and decuplet baryons are scalar and axial-vector diquarks, primarily because the associated mass-scales are smaller than the masses of these baryons \cite{cjbsep,marisdq} and the positive parity of the correlations matches that of the baryons.  Both scalar and axial-vector diquarks provide attraction in the Faddeev equation; e.g., a scalar diquark alone provides for a bound octet baryon and including axial-vector correlations reduces that baryon's mass.  

With the retention of axial-vector diquark correlations a quantitative description of baryon properties is attainable \cite{oettelfe}.  However, that possibility necessitates the incorporation of pseudoscalar meson loop  contributions because a credible description of baryon properties is otherwise problematic.  Such effects contribute materially: to baryon masses \cite{hechtfe,cbm,wright}; and charge and magnetic radii, and magnetic moments \cite{radiiCh,young}.  

In Sec.\,\ref{Sec:Faddeev} we recapitulate on those aspects of the Poincar\'e covariant Faddeev equations for the nucleon and $\Delta$ that are important to our analysis, and report the solutions.  Section~\ref{Ncurrent} describes the formulation of a Ward-Takahashi identity preserving current that is appropriate to a nucleon represented by a solution of the Faddeev equation.  With the necessary elements thus specified, the nucleons' electromagnetic form factors are presented and discussed in Sec.\,\ref{FFs}.
%: this section is self-contained.  
Section~\ref{epilogue} is an epilogue.

\section{Covariant Faddeev Equation}
\label{Sec:Faddeev}
The properties of light pseudoscalar and vector mesons are described well by
a rainbow-ladder truncation of QCD's Dyson-Schwinger equations (DSEs) \cite{marisrev}, and the calculation of baryon properties using the solution of a Poincar\'e covariant Faddeev equation is a desirable extension of that approach.  For quarks in the fundamental representation of colour-$SU(3)$:
\begin{equation} 
3_c \otimes 3_c \otimes 3_c = (\bar 3_c \oplus 6_c) \otimes 3_c = 1_c \oplus 
8_c^\prime \oplus 8_c \oplus 10_c\,,
\end{equation} 
and hence any two quarks in a colour-singlet three-quark bound state must constitute a relative colour-antitriplet.  This enables the derivation of a Faddeev equation for the bound state contribution to the three quark scattering kernel because the same kernel that describes mesons so well is also attractive for quark-quark scattering in the colour-$\bar 3$ channel.  

In this truncation of the three-body problem the interactions between two selected quarks are added to yield a quark-quark scattering matrix, which is then approximated as a sum over all possible diquark pseudoparticle terms  \cite{cjbfe}: Dirac-scalar $+$ -axial-vector $+ [\ldots]$.  The Faddeev equation thus obtained describes the baryon as a composite of a dressed-quark and nonpointlike diquark with an iterated exchange of roles between the bystander and diquark-participant quarks.  The baryon is consequently represented by a Faddeev amplitude: 
\begin{equation} 
\Psi = \Psi_1 + \Psi_2 + \Psi_3 \,, 
\end{equation} 
where the subscript identifies the bystander quark and, e.g., $\Psi_{1,2}$ are obtained from $\Psi_3$ by a correlated, cyclic permutation of all the quark 
labels.  The Faddeev amplitude thus obtained does not contain explicit three-body correlations.

\subsection{{\it Ans\"atze} for the Nucleon and $\Delta$}
\label{ANDelta}
We employ the simplest realistic representation of the Faddeev amplitudes for the nucleon and $\Delta$.  The spin- and isospin-$1/2$ nucleon is a sum of scalar and axial-vector diquark correlations:
\begin{equation} 
\label{Psi} \Psi_3(p_i,\alpha_i,\tau_i) = {\cal N}_3^{0^+} + {\cal N}_3^{1^+}, 
\end{equation} 
with $(p_i,\alpha_i,\tau_i)$ the momentum, spin and isospin labels of the 
quarks constituting the bound state, and $P=p_1+p_2+p_3$ the system's total momentum.  NB.\ We assume isospin symmetry of the strong interaction throughout; i.e., the $u$- and $d$-quarks are indistinguishable but for their electric charge.  Since it is not possible to combine an isospin-$0$ diquark with an isospin-$1/2$ quark to obtain isospin-$3/2$, the spin- and isospin-$3/2$ $\Delta$ contains only an axial-vector diquark component
\begin{equation}
\label{PsiD} \Psi^\Delta_3(p_i,\alpha_i,\tau_i) = {\cal D}_3^{1^+}.
\end{equation} 
%A term representing the quark coupled to a scalar diquark with a single unit of angular momentum is forbidden by Fermi statistics when isospin symmetry is assumed.

The scalar diquark piece in Eq.\,(\ref{Psi}) is 
\begin{eqnarray} 
{\cal N}_3^{0^+}(p_i,\alpha_i,\tau_i)&=& [\Gamma^{0^+}(\sfrac{1}{2}p_{[12]};K)]_{\alpha_1 
\alpha_2}^{\tau_1 \tau_2}\, \Delta^{0^+}(K) \,[{\cal S}(\ell;P) u(P)]_{\alpha_3}^{\tau_3}\,,% 
\label{calS} 
\end{eqnarray} 
where:\footnote{The metric we employ is described in \protect\ref{App:EM}~Euclidean Conventions.} the spinor satisfies 
\begin{equation}
(i\gamma\cdot P + M)\, u(P) =0= \bar u(P)\, (i\gamma\cdot P + M)\,,
\end{equation}
with $M$ the mass obtained by solving the Faddeev equation, and it is also a
spinor in isospin space with $\varphi_+= {\rm col}(1,0)$ for the proton and
$\varphi_-= {\rm col}(0,1)$ for the neutron; $K= p_1+p_2=: p_{\{12\}}$,
$p_{[12]}= p_1 - p_2$, $\ell := (-p_{\{12\}} + 2 p_3)/3$; $\Delta^{0^+}$
is a pseudoparticle propagator for the scalar diquark formed from quarks $1$
and $2$, and $\Gamma^{0^+}\!$ is a Bethe-Salpeter-like amplitude describing
their relative momentum correlation; and ${\cal S}$, a $4\times 4$ Dirac
matrix, describes the relative quark-diquark momentum correlation.  (${\cal
S}$, $\Gamma^{0^+}$ and $\Delta^{0^+}$ are discussed in Sec.\,\ref{completing}.)  The colour antisymmetry of $\Psi_3$ is implicit in $\Gamma^{J^P}\!\!$, with the 
Levi-Civita tensor, $\epsilon_{c_1 c_2 c_3}$, expressed via the antisymmetric 
Gell-Mann matrices; viz., defining 
\begin{equation} 
\{H^1=i\lambda^7,H^2=-i\lambda^5,H^3=i\lambda^2\}\,, 
\end{equation} 
then $\epsilon_{c_1 c_2 c_3}= (H^{c_3})_{c_1 c_2}$.  [See 
Eqs.\,(\ref{Gamma0p}), (\ref{Gamma1p}).]

The axial-vector component in Eq.\,(\ref{Psi}) is
\begin{eqnarray} 
{\cal N}^{1^+}(p_i,\alpha_i,\tau_i) & =&  [{\tt t}^i\,\Gamma_\mu^{1^+}(\sfrac{1}{2}p_{[12]};K)]_{\alpha_1 
\alpha_2}^{\tau_1 \tau_2}\,\Delta_{\mu\nu}^{1^+}(K)\, 
[{\cal A}^{i}_\nu(\ell;P) u(P)]_{\alpha_3}^{\tau_3}\,,
\label{calA} 
\end{eqnarray} 
where the symmetric isospin-triplet matrices are 
\begin{equation} 
{\tt t}^+ = \frac{1}{\surd 2}(\tau^0+\tau^3) \,,\; 
{\tt t}^0 = \tau^1\,,\; 
{\tt t}^- = \frac{1}{\surd 2}(\tau^0-\tau^3)\,, 
\end{equation} 
with $(\tau^0)_{ij}=\delta_{ij}$ and $\tau^{1,3}$ the usual Pauli matrices, 
and the other elements in Eq.\,(\ref{calA}) are straightforward generalisations of those in Eq.\,(\ref{calS}). 

The general form of the Faddeev amplitude for the spin- and isospin-$3/2$ $\Delta$ is complicated.  However, isospin symmetry means one can focus on the $\Delta^{++}$ with it's simple flavour structure, because all the charge states are degenerate, and consider 
\begin{equation}
{\cal D}_3^{1^+}= [{\tt t}^+ \Gamma^{1^+}_\mu(\sfrac{1}{2}p_{[12]};K)]_{\alpha_1 \alpha_2}^{\tau_1 \tau_2}
\, \Delta_{\mu\nu}^{1^+}(K) \, [{\cal D}_{\nu\rho}(\ell;P)u_\rho(P)\, \varphi_+]_{\alpha_3}^{\tau_3}\,, \label{DeltaAmpA} 
\end{equation} 
where $u_\rho(P)$ is a Rarita-Schwinger spinor, Eq.\,(\ref{rarita}).

The general forms of the matrices ${\cal S}(l;P)$, ${\cal A}^i_\nu(l;P)$ and ${\cal D}_{\nu\rho}(\ell;P)$, which describe the momentum space correlation between the quark and diquark in the nucleon and the $\Delta$, respectively, are described in Ref.\,\cite{oettelfe}.  The requirement that ${\cal S}(l;P)$ represent a positive energy nucleon; namely, that it be an eigenfunction of $\Lambda_+(P)$, Eq.\,(\ref{Lplus}), entails
\begin{equation}
\label{Sexp} 
{\cal S}(\ell;P) = s_1(\ell;P)\,I_{\rm D} + \left(i\gamma\cdot \hat\ell - \hat\ell \cdot \hat P\, I_{\rm D}\right)\,s_2(\ell;P)\,, 
\end{equation} 
where $(I_{\rm D})_{rs}= \delta_{rs}$, $\hat l^2=1$, $\hat P^2= - 1$.  In the nucleon rest frame, $s_{1,2}$ describe, respectively, the upper, lower component of the bound-state nucleon's spinor.  Placing the same constraint on the axial-vector component, one has
\begin{equation}
\label{Aexp}
 {\cal A}^i_\nu(l;P) = \sum_{n=1}^6 \, p_n^i(l;P)\,\gamma_5\,A^n_{\nu}(l;P)\,,\; i=+,0,-\,,
\end{equation}
where ($ \hat l^\perp_\nu = \hat l_\nu + \hat l\cdot\hat P\, \hat P_\nu$, $ \gamma^\perp_\nu = \gamma_\nu + \gamma\cdot\hat P\, \hat P_\nu$)
\begin{equation}
\begin{array}{lll}
A^1_\nu= \gamma\cdot \hat l^\perp\, \hat P_\nu \,,\; &
A^2_\nu= -i \hat P_\nu \,,\; &
A^3_\nu= \gamma\cdot\hat l^\perp\,\hat l^\perp\,,\\
A^4_\nu= i \,\hat l_\mu^\perp\,,\; &
A^5_\nu= \gamma^\perp_\nu - A^3_\nu \,,\; &
A^6_\nu= i \gamma^\perp_\nu \gamma\cdot\hat l^\perp - A^4_\nu\,.
\end{array}
\end{equation}
Finally, requiring also that ${\cal D}_{\nu\rho}(\ell;P)$ be an eigenfunction of $\Lambda_+(P)$, one obtains
\begin{equation}
{\cal D}_{\nu\rho}(\ell;P) = {\cal S}^\Delta(l;P) \, \delta_{\nu\rho} + \gamma_5{\cal A}_\nu^\Delta(l;P) \,l^\perp_\rho \,,
\end{equation}
with ${\cal S}^\Delta$ and ${\cal A}^\Delta_\nu$ given by obvious analogues of Eqs.\,(\ref{Sexp}) and (\ref{Aexp}), respectively.

One can now write the Faddeev equation satisfied by $\Psi_3$ as
\begin{equation} 
 \left[ \begin{array}{r} 
{\cal S}(k;P)\, u(P)\\ 
{\cal A}^i_\mu(k;P)\, u(P) 
\end{array}\right]\\ 
 = -4\,\int\frac{d^4\ell}{(2\pi)^4}\,{\cal M}(k,\ell;P) 
\left[ 
\begin{array}{r} 
{\cal S}(\ell;P)\, u(P)\\ 
{\cal A}^j_\nu(\ell;P)\, u(P) 
\end{array}\right], 
\label{FEone} 
\end{equation} 
where one factor of ``2'' appears because $\Psi_3$ is coupled symmetrically to $\Psi_1$ and $\Psi_2$, and the necessary colour contraction has been evaluated: $(H^a)_{bc} (H^a)_{cb^\prime}=-2 \,\delta_{bb^\prime}$.  The kernel in Eq.~(\ref{FEone}) is 
\begin{equation} 
\label{calM} {\cal M}(k,\ell;P) = \left[\begin{array}{cc} 
{\cal M}_{00} & ({\cal M}_{01})^j_\nu \\ 
({\cal M}_{10})^i_\mu & ({\cal M}_{11})^{ij}_{\mu\nu}\rule{0mm}{3ex} 
\end{array} 
\right] 
\end{equation} 
with 
\begin{equation} 
 {\cal M}_{00} = \Gamma^{0^+}\!(k_q-\ell_{qq}/2;\ell_{qq})\, 
S^{\rm T}(\ell_{qq}-k_q) \,\bar\Gamma^{0^+}\!(\ell_q-k_{qq}/2;-k_{qq})\, 
S(\ell_q)\,\Delta^{0^+}(\ell_{qq}) \,, 
\end{equation} 
where:\footnote{This choice is explained by Eq.\,(\protect\ref{etavalue}) and the discussion thereabout.} $\ell_q=\ell+P/3$, $k_q=k+P/3$, $\ell_{qq}=-\ell+ 2P/3$, 
$k_{qq}=-k+2P/3$ and the superscript ``T'' denotes matrix transpose; and
\begin{eqnarray}
\nonumber
\lefteqn{({\cal M}_{01})^j_\nu= {\tt t}^j \,
\Gamma_\mu^{1^+}\!(k_q-\ell_{qq}/2;\ell_{qq})} \\
&& \times 
S^{\rm T}(\ell_{qq}-k_q)\,\bar\Gamma^{0^+}\!(\ell_q-k_{qq}/2;-k_{qq})\, 
S(\ell_q)\,\Delta^{1^+}_{\mu\nu}(\ell_{qq}) \,, \label{calM01} \\ 
\nonumber \lefteqn{({\cal M}_{10})^i_\mu = 
\Gamma^{0^+}\!(k_q-\ell_{qq}/2;\ell_{qq})\, 
}\\ 
&&\times S^{\rm T}(\ell_{qq}-k_q)\,{\tt t}^i\, \bar\Gamma_\mu^{1^+}\!(\ell_q-k_{qq}/2;-k_{qq})\, 
S(\ell_q)\,\Delta^{0^+}(\ell_{qq}) \,,\\ 
\nonumber \lefteqn{({\cal M}_{11})^{ij}_{\mu\nu} = {\tt t}^j\, 
\Gamma_\rho^{1^+}\!(k_q-\ell_{qq}/2;\ell_{qq})}\\ 
&&\times \, S^{\rm T}(\ell_{qq}-k_q)\,{\tt t}^i\, \bar\Gamma^{1^+}_\mu\!(\ell_q-k_{qq}/2;-k_{qq})\, 
S(\ell_q)\,\Delta^{1^+}_{\rho\nu}(\ell_{qq}) \,. \label{calM11} 
\end{eqnarray} 

It is illuminating to note that $u(P)$ in Eq.\,(\ref{FEone}) is a normalised average of $\varphi_\pm$ so that, e.g., the proton equation is obtained by projection on the left with $\varphi^\dagger_+$.  To illustrate this we note that Eq.\,(\ref{calM01}) generates an isospin coupling 
between $u(P)_{\varphi_+}$ on the left-hand-side (l.h.s.) of Eq.\,(\ref{FEone}) and, on the r.h.s., 
\begin{equation} 
\sqrt 2\,{\cal A}^+_\nu\, u(P)_{\varphi^-} - {\cal A}^0_\nu 
\,u(P)_{\varphi_+}\,. 
\end{equation} 
This is just the Clebsch-Gordon coupling of isospin-$1\oplus\,$isospin-$\frac{1}{2}$ to total isospin-$\frac{1}{2}$ and 
means that the scalar diquark amplitude in the proton, $(ud)_{0^+}\,u$, is 
coupled to itself {\it and} the linear combination: 
\begin{equation} 
\sqrt 2\, (uu)_{1^+}\, d - (ud)_{1^+} \, u\,. 
\end{equation} 
Similar statements are obviously true of the spin couplings.

The $\Delta$'s Faddeev equation is
\begin{eqnarray} 
{\cal D}_{\lambda\rho}(k;P)\,u_\rho(P) & = & 4\int\frac{d^4\ell}{(2\pi)^4}\,{\cal 
M}^\Delta_{\lambda\mu}(k,\ell;P) \,{\cal D}_{\mu\sigma}(\ell;P)\,u_\sigma(P)\,, \label{FEDelta} 
\end{eqnarray} 
with
\begin{equation}
{\cal M}^\Delta_{\lambda\mu} = {\tt t}^+ 
\Gamma_\sigma^{1^+}\!(k_q-\ell_{qq}/2;\ell_{qq})\,
 S^{\rm T}\!(\ell_{qq}-k_q)\, {\tt t}^+\bar\Gamma^{1^+}_\lambda\!(\ell_q-k_{qq}/2;-k_{qq})\, 
S(\ell_q)\,\Delta^{1^+}_{\sigma\mu}\!(\ell_{qq}). 
\end{equation}

\subsection{Completing the Faddeev Equation Kernels}
\label{completing}
To complete the Faddeev equations, Eqs.\,(\ref{FEone}) \& (\ref{FEDelta}), one must specify the dressed-quark propagator, the diquark Bethe-Salpeter amplitudes and the diquark propagators that appear in the kernels.
\subsubsection{Dressed-quark propagator} 
\label{subsubsec:S} 
The dressed-quark propagator can be obtained from QCD's gap equation and the general form of the solution is 
\begin{eqnarray} 
S(p) & = & -i \gamma\cdot p\, \sigma_V(p^2) + \sigma_S(p^2) = 1/[i\gamma\cdot p\, A(p^2) + B(p^2)]\,.\label{SpAB} 
\end{eqnarray}  
It is a longstanding prediction of DSE studies in QCD that the wave function renormalisation and dressed-quark mass: 
\begin{equation} 
\label{ZMdef}
Z(p^2)=1/A(p^2)\,,\;M(p^2)=B(p^2)/A(p^2)\,, 
\end{equation} 
respectively, receive strong momentum-dependent corrections at infrared momenta \cite{lane,politzer,cdragw}: $Z(p^2)$ is suppressed and $M(p^2)$ enhanced.  The prediction was confirmed in recent simulations \cite{bowman} of lattice-regularised quenched-QCD,\footnote{For three light flavours the effects of unquenching appear to be small \protect\cite{fischer}.} and the conditions under which pointwise agreement between DSE results and lattice simulations may be obtained have been explored \cite{bhagwat,maris,bhagwat2}. The enhancement of $M(p^2)$ is central to the appearance of a constituent-quark mass-scale and an existential prerequisite for Goldstone modes.  The mass function evolves with increasing $p^2$ to reproduce the asymptotic behaviour familiar from perturbative analyses, and that behaviour is unambiguously evident for $p^2 \gtrsim 10\,$GeV$^2$ \cite{mishasvy,wrightgap}. 

The importance of this infrared dressing has long been emphasised in DSE studies of hadron phenomena \cite{cdrpion} and, while numerical solutions of the quark DSE are now readily obtained, the utility of an algebraic form
for $S(p)$ when calculations require the evaluation of numerous
multidimensional integrals is self-evident.  An efficacious parametrisation 
of $S(p)$, which exhibits the features described above, has been used 
extensively in hadron studies \cite{marisrev,bastirev,alkoferrev}.  It is expressed via
\begin{eqnarray} 
\bar\sigma_S(x) & =&  2\,\bar m \,{\cal F}(2 (x+\bar m^2)) + {\cal
F}(b_1 x) \,{\cal F}(b_3 x) \,  
\left[b_0 + b_2 {\cal F}(\epsilon x)\right]\,,\label{ssm} \\ 
\label{svm} \bar\sigma_V(x) & = & \frac{1}{x+\bar m^2}\, \left[ 1 - {\cal F}(2 (x+\bar m^2))\right]\,, 
\end{eqnarray}
with $x=p^2/\lambda^2$, $\bar m$ = $m/\lambda$, 
\begin{equation}
\label{defcalF}
{\cal F}(x)= \frac{1-\mbox{\rm e}^{-x}}{x}  \,, 
\end{equation}
$\bar\sigma_S(x) = \lambda\,\sigma_S(p^2)$ and $\bar\sigma_V(x) =
\lambda^2\,\sigma_V(p^2)$.  The mass-scale, $\lambda=0.566\,$GeV, and
parameter values\footnote{$\epsilon=10^{-4}$ in Eq.\ (\ref{ssm}) acts only to
decouple the large- and intermediate-$p^2$ domains.}
\begin{equation} 
\label{tableA} 
\begin{array}{ccccc} 
   \bar m& b_0 & b_1 & b_2 & b_3 \\\hline 
   0.00897 & 0.131 & 2.90 & 0.603 & 0.185 
\end{array}\;, 
\end{equation} 
were fixed in a least-squares fit to light-meson observables \cite{mark}.  The dimensionless $u=d$ current-quark mass in Eq.~(\ref{tableA}) corresponds to
\begin{equation} 
m_{u,d}=5.1\,{\rm MeV}\,. 
\end{equation} 

The parametrisation yields a Euclidean constituent-quark mass
\begin{equation} 
\label{MEq} M_{u,d}^E = 0.33\,{\rm GeV}, 
\end{equation} 
defined as the solution of $p^2=M^2(p^2)$ \cite{mr97}, whose magnitude is
typical of that employed in constituent-quark models \cite{simon}.  In Ref.\,\cite{mark}, $m_s=25\,m_{u,d}$ and $M_{s}^E = 0.49\,{\rm GeV}$.  It is generally true that $M_{s}^E- M_{u,d}^E \gtrsim \hat m_s - \hat m_{u,d}$, where $\hat m$ denotes the renomalisation point independent current-quark mass.  The constituent-quark mass is an expression of dynamical chiral symmetry breaking, as is the vacuum quark condensate\footnote{The condensate is calculated directly from its gauge invariant definition \cite{mrt98} after making allowance for the fact that Eqs.\,(\ref{ssm}) \& (\ref{svm}) yield a chiral-limit quark mass function with anomalous dimension $\gamma_m = 1$.  This omission of the additional $\ln( p^2/\Lambda_{\rm QCD}^2)$-suppression that is characteristic of QCD is merely a practical simplification.}  ($\Lambda_{\rm QCD}=0.2\,$GeV)
\begin{equation} 
-\langle \bar qq \rangle_0^{1\,{\rm GeV}^2} = \lambda^3\,\frac{3}{4\pi^2}\, 
\frac{b_0}{b_1\,b_3}\,\ln\frac{1\,{\rm GeV}^2}{\Lambda_{\rm QCD}^2} = 
(0.221\,{\rm GeV})^3\,. 
\end{equation}  

Motivated by model DSE studies \cite{entire1,entire2}, Eqs.~(\ref{ssm}) \& (\ref{svm}) express the dressed-quark propagator as an entire function.  Hence $S(p)$ does not have a Lehmann representation, which is a sufficient condition for confinement.\footnote{It is a sufficient condition for confinement 
because of the associated violation of reflection positivity.  This notion may be traced from Refs.\,\cite{entire1,entire2,stingl,krein} and is reviewed in Refs.\,\cite{cdragw,bastirev,alkoferrev}.}  Employing an entire function, whose form is only constrained through the calculation of spacelike observables, can lead to model artefacts when it is employed directly to calculate observables involving large timelike momenta \cite{ahlig}.  An improved parametrisation is therefore being sought.  Nevertheless, difficulties are not encountered for moderate timelike momenta, and on the domain of the complex plane explored in the present calculation the integral support provided by an equally effective alternative cannot differ significantly from that of this parametrisation. 

\subsubsection{Diquark Bethe-Salpeter amplitudes}
\label{qqBSA}
The rainbow-ladder DSE truncation yields asymptotic diquark states in the strong interaction spectrum.  Such states are not observed and their appearance is an artefact of the truncation.  Higher order terms in the 
quark-quark scattering kernel, whose analogue in the quark-antiquark channel do not much affect the properties of vector and flavour non-singlet pseudoscalar mesons, ensure that QCD's quark-quark scattering matrix does not exhibit singularities which correspond to asymptotic diquark states~\cite{mandarvertex}.  Nevertheless, studies with kernels that do not produce diquark bound states, do support a physical interpretation of the masses, $m_{(qq)_{J^P}}$, obtained using the rainbow-ladder truncation: the quantity $l_{(qq)_{J^P}}=1/m_{(qq)_{J^P}}$ may be interpreted as a range over which the diquark correlation can persist inside a baryon.  These observations motivate the {\it Ansatz} for the quark-quark scattering matrix that is employed in deriving the Faddeev equation: 
\begin{equation} 
[M_{qq}(k,q;K)]_{rs}^{tu} = \sum_{J^P=0^+,1^+,\ldots} \bar\Gamma^{J^P}\!(k;-K)\, \Delta^{J^P}\!(K) \, \Gamma^{J^P}\!(q;K)\,. \label{AnsatzMqq} 
\end{equation}  

One practical means of specifying the $\Gamma^{J^P}\!\!$ in Eq.\,(\ref{AnsatzMqq}) is to employ the solutions of a rainbow-ladder quark-quark Bethe-Salpeter equation (BSE).  Using the properties of the Gell-Mann matrices one finds easily that $\Gamma^{J^P}_C:= \Gamma^{J^P}C^\dagger$ satisfies exactly the same equation as the $J^{-P}$ colour-singlet meson {\it but} for a halving of the coupling strength \cite{regdq}.  This makes clear that the interaction in the ${\bar 3_c}$ $(qq)$ channel is strong and attractive.\footnote{The same analysis shows the interaction to be strong and repulsive in the ${6_c}$ $(qq)$ channel.}  Moreover, it follows as a feature of the rainbow-ladder truncation that, independent of the specific form of a model's interaction, the calculated masses satisfy 
\begin{equation}
m_{(qq)_{J^P}} > m_{(\bar q q)_{J^{-P}}}\,.
\end{equation}
This is a useful guide for all but scalar diquark correlations because the partnered mesons in that case are pseudoscalars, whose ground state masses are constrained to be small by Goldstone's theorem and which therefore provide a weak lower bound.  For the correlations relevant herein, models typically give masses (in GeV) \cite{cjbsep,marisdq}:
\begin{equation}
\label{diquarkmass}
m_{(ud)_{0^+}} = 0.74 - 0.82 \,,\; m_{(uu)_{1^+}}=m_{(ud)_{1^+}}=m_{(dd)_{1^+}}=0.95 - 1.02\,.
\end{equation}
%When one of the light-quarks is replaced by a $s$-quark these masses increase by $\lesssim 2\, (M^E_s-M^E_u)$.
Such values are confirmed by results obtained in simulations of quenched lattice-QCD \cite{hess}.  Reference~\cite{marisdqff} also evaluates charge radii for the scalar diquarks: $r_{(ud)_{0^+}}\approx 1.1\,r_\pi$; $r_{(ds)_{0^+}}\approx r_{K^+}$; and $r_{(us)_{0^+}}\approx 1.3\,$-$1.4\,r_{K^+}$.  The Bethe-Salpeter amplitude is canonically normalised via: 
\begin{eqnarray}
\label{BSEnorm} 
2 \,K_\mu & = & 
\left[ \frac{\partial}{\partial Q_\mu} \Pi(K,Q) \right]_{Q=K}^{{K^2=-m_{J^P}^2}},\\
\Pi(K,Q) & = & tr\!\! \int\!\! 
\frac{d^4 q}{(2\pi)^4}\, \bar\Gamma(q;-K) \, S(q+Q/2) \, \Gamma(q;K) \, S^{\rm T}(-q+Q/2) .
\end{eqnarray}

A solution of the BSE equation requires a simultaneous solution of the quark-DSE \cite{marisdq}.  However, since we have already chosen to simplify the calculations by parametrising $S(p)$, we 
follow Ref.~\cite{hechtfe} and also employ that expedient with $\Gamma^{J^P}\!$, using the following one-parameter forms: 
\begin{eqnarray} 
\label{Gamma0p} \Gamma^{0^+}(k;K) &=& \frac{1}{{\cal N}^{0^+}} \, 
H^a\,C i\gamma_5\, i\tau_2\, {\cal F}(k^2/\omega_{0^+}^2) \,, \\ 
\label{Gamma1p} {\tt t}^i \Gamma^{1^+}_\mu (k;K) &=& \frac{1}{{\cal N}^{1^+}}\, 
H^a\,i\gamma_\mu C\,{\tt t}^i\, {\cal F}(k^2/\omega_{1^+}^2)\,, 
\end{eqnarray} 
with the normalisation, ${\cal N}^{J^P}\!$, fixed by Eq.~(\ref{BSEnorm}). These {\it Ans\"atze} retain only that single Dirac-amplitude which would represent a point particle with the given quantum numbers in a local Lagrangian density: they are usually the dominant amplitudes in a solution of the rainbow-ladder BSE for the lowest mass $J^P$ diquarks \cite{cjbsep,marisdq} and mesons \cite{mr97,pieter,pieterpion}. 

\subsubsection{Diquark propagators}
\label{qqprop}
Solving for the quark-quark scattering matrix using the rainbow-ladder truncation yields free particle propagators for $\Delta^{J^P}$ in 
Eq.~(\ref{AnsatzMqq}).  As already noted, however, higher order contributions 
remedy that defect, eliminating asymptotic diquark states from the spectrum. It is apparent in Ref.~\cite{mandarvertex} that the attendant modification of 
$\Delta^{J^P}$ can be modelled efficiently by simple functions that are 
free-particle-like at spacelike momenta but pole-free on the timelike axis. 
Hence we employ 
\begin{eqnarray} 
\Delta^{0^+}(K) & = & \frac{1}{m_{0^+}^2}\,{\cal F}(K^2/\omega_{0^+}^2)\,,\\ 
\Delta^{1^+}_{\mu\nu}(K) & = & 
\left(\delta_{\mu\nu} + \frac{K_\mu K_\nu}{m_{1^+}^2}\right) \, \frac{1}{m_{1^+}^2}\, {\cal F}(K^2/\omega_{1^+}^2) \,,
\end{eqnarray} 
where the two parameters $m_{J^P}$ are diquark pseudoparticle masses and 
$\omega_{J^P}$ are widths characterising $\Gamma^{J^P}\!$.  Herein we require additionally that
\begin{equation}
\label{DQPropConstr}
\left. \frac{d}{d K^2}\,\left(\frac{1}{m_{J^P}^2}\,{\cal F}(K^2/\omega_{J^P}^2)\right)^{-1} \right|_{K^2=0}\! = 1 \; \Rightarrow \; \omega_{J^P}^2 = \sfrac{1}{2}\,m_{J^P}^2\,,
\end{equation} 
which is a normalisation that accentuates the free-particle-like propagation characteristics of the diquarks {\it within} the hadron. 

\subsection{Nucleon and $\Delta$ Masses}
\label{NDmasses}
All elements of the Faddeev equations, Eqs.\,(\ref{FEone}) \& (\ref{FEDelta}), are now completely specified.  We solve the equations via the method described in Ref.\,\cite{oettelcomp}.  The masses of the scalar and axial-vector diquarks are the only variable parameters.  The axial-vector mass is chosen so as to obtain a desired mass for the $\Delta$, and the scalar mass is subsequently set by requiring a particular nucleon mass.  

\begin{table}[b]
\begin{center}
\caption{\label{ParaFix} Mass-scale parameters (in GeV) for the scalar and axial-vector diquark correlations, fixed by fitting nucleon and $\Delta$ masses: for Set~A, a fit to the actual masses was required; whereas for Set~B the fitted mass was offset to allow for ``pion cloud'' contributions \protect\cite{hechtfe}.  We also list $\omega_{J^{P}}= \sfrac{1}{\surd 2}m_{J^{P}}$, which is the width-parameter in the $(qq)_{J^P}$ Bethe-Salpeter amplitude, Eqs.\,(\protect\ref{Gamma0p}) \& (\protect\ref{Gamma1p}):  its inverse is a gauge of the diquark's matter radius.}
\begin{tabular*}{1.0\textwidth}{
c@{\extracolsep{0ptplus1fil}}c@{\extracolsep{0ptplus1fil}}c@{\extracolsep{0ptplus1fil}}|c@{\extracolsep{0ptplus1fil}} c@{\extracolsep{0ptplus1fil}}|c@{\extracolsep{0ptplus1fil}}c@{\extracolsep{0ptplus1fil}}}
\hline
set & $M_N$ & $M_{\Delta}$~ & $m_{0^{+}}$ & $m_{1^{+}}$~ &
$\omega_{0^{+}} $ & $\omega_{1^{+}}$ \\
\hline
A & 0.94 & 1.23~ & 0.63 & 0.84~ & 0.44=1/(0.45\,{\rm fm}) & 0.59=1/(0.33\,{\rm fm}) \\
B & 1.18 & 1.33~ & 0.79 & 0.89~ & 0.56=1/(0.35\,{\rm fm}) & 0.63=1/(0.31\,{\rm fm}) \\
\hline
\end{tabular*}
\end{center}
\end{table}

Two parameter sets are presented in Table~\ref{ParaFix}.  We obtained Set~A by requiring a precise fit to the experimental nucleon and $\Delta$ masses.  It has long been known that this is possible; e.g., Ref.\,\cite{oettelfe} reports octet and decuplet baryon masses in which the rms deviation between the calculated mass and experiment is only $2$\%.  However, it is also known that such an outcome is undesirable because, e.g., studies using the cloudy bag model \cite{cbm} indicate that the nucleon's mass is reduced by as much as $\delta M_N = -300$ to $-400\,$MeV through pion self-energy corrections \cite{bruceCBM}.  Furthermore, a perturbative study, using the Faddeev equation, of the mass shift induced by pion exchange between the quark and diquark constituents of the nucleon obtains $\delta M_N = -150$ to $-300\,$MeV~\cite{ishii}.  We are thus led to Set~B, which was obtained by fitting to nucleon and $\Delta$ masses that are inflated so as to allow for the additional attractive contribution from the pion cloud \cite{hechtfe}. 

It is apparent in Table~\ref{ParaFix} that a baryon's mass increases with increasing diquark mass, and the fitted diquark mass-scales are commensurate with the anticipated values, cf.\ Eq.\,(\ref{diquarkmass}), with Set~B in better accord.  If coupling to the axial-vector diquark channel is omitted from Eq.\,(\ref{FEone}), then $M_N^{\rm Set\,A} = 1.15\,$GeV and $M_N^{\rm Set\,B} = 1.46\,$GeV.  It is thus clear that axial-vector diquark correlations provide significant attraction in the nucleon.  Of course, using our Faddeev equation, the $\Delta$ does not exist without axial-vector correlations.  In Set~B the amount of attraction provided by axial-vector correlations must be matched by that provided by the pion cloud.  This highlights the constructive interference between the contribution of these two effects to a baryons' mass.  It is related and noteworthy that $m_{1^+}-m_{0^+}$ is only a reasonable approximation to $M_\Delta - M_N=0.29\,$GeV when pion cloud effects are ignored: Set~A, $m_{1^+}-m_{0^+}=0.21\,$GeV cf.\ Set~B, $m_{1^+}-m_{0^+}=0.10\,$GeV.  Plainly, understanding the $N$-$\Delta$ mass splitting requires more than merely reckoning the mass-scales of constituent degrees of freedom.  

\section{Nucleon-Photon Vertex}
\label{Ncurrent}
The nucleon's electromagnetic current is
\begin{eqnarray}
\label{Jnucleon}
J_\mu(P^\prime,P) & = & ie\,\bar u(P^\prime)\, \Lambda_\mu(q,P) \,u(P)\,, \\
& = &  i e \,\bar u(P^\prime)\,\left( \gamma_\mu F_1(Q^2) +
\frac{1}{2M}\, \sigma_{\mu\nu}\,Q_\nu\,F_2(Q^2)\right) u(P)\,,
\label{JnucleonB}
\end{eqnarray}
where $P$ ($P^\prime$) is the momentum of the incoming (outgoing) nucleon, $Q= P^\prime - P$, and $F_1$ and $F_2$ are, respectively, the Dirac and Pauli form factors.  They are the primary calculated quantities, from which one obtains the nucleon's electric and magnetic form factors
\begin{equation}
\label{GEpeq}
G_E(Q^2)  =  F_1(Q^2) - \frac{Q^2}{4 M^2} F_2(Q^2)\,,\; 
G_M(Q^2)  =  F_1(Q^2) + F_2(Q^2)\,.
\end{equation}

\begin{figure}[t]
\begin{minipage}[t]{\textwidth}
\begin{minipage}[t]{0.45\textwidth}
\leftline{\includegraphics[width=0.90\textwidth]{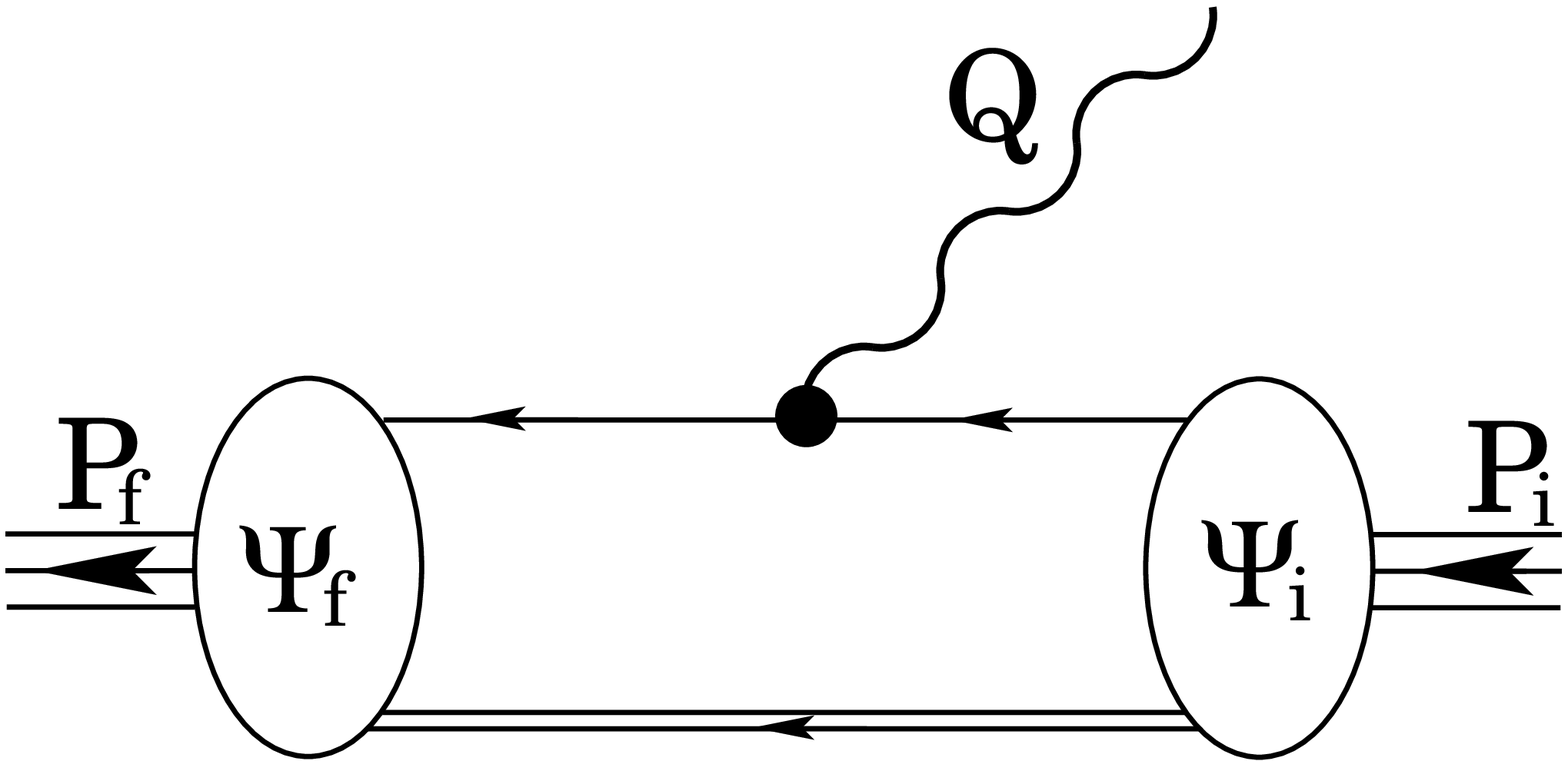}}
\end{minipage}
\begin{minipage}[t]{0.45\textwidth}
\rightline{\includegraphics[width=0.90\textwidth]{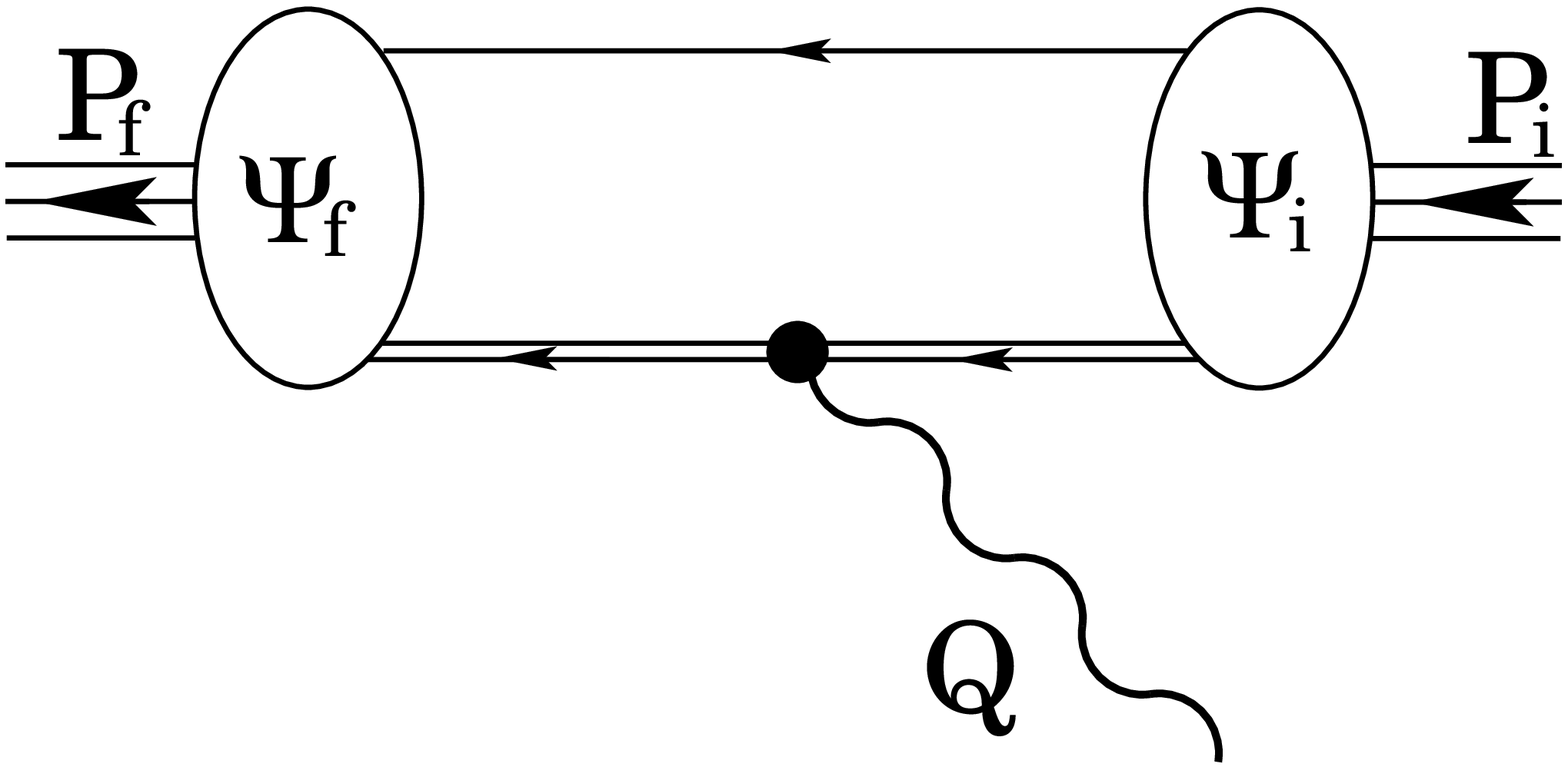}}
\end{minipage}\vspace*{3ex}

\begin{minipage}[t]{0.45\textwidth}
\leftline{\includegraphics[width=0.90\textwidth]{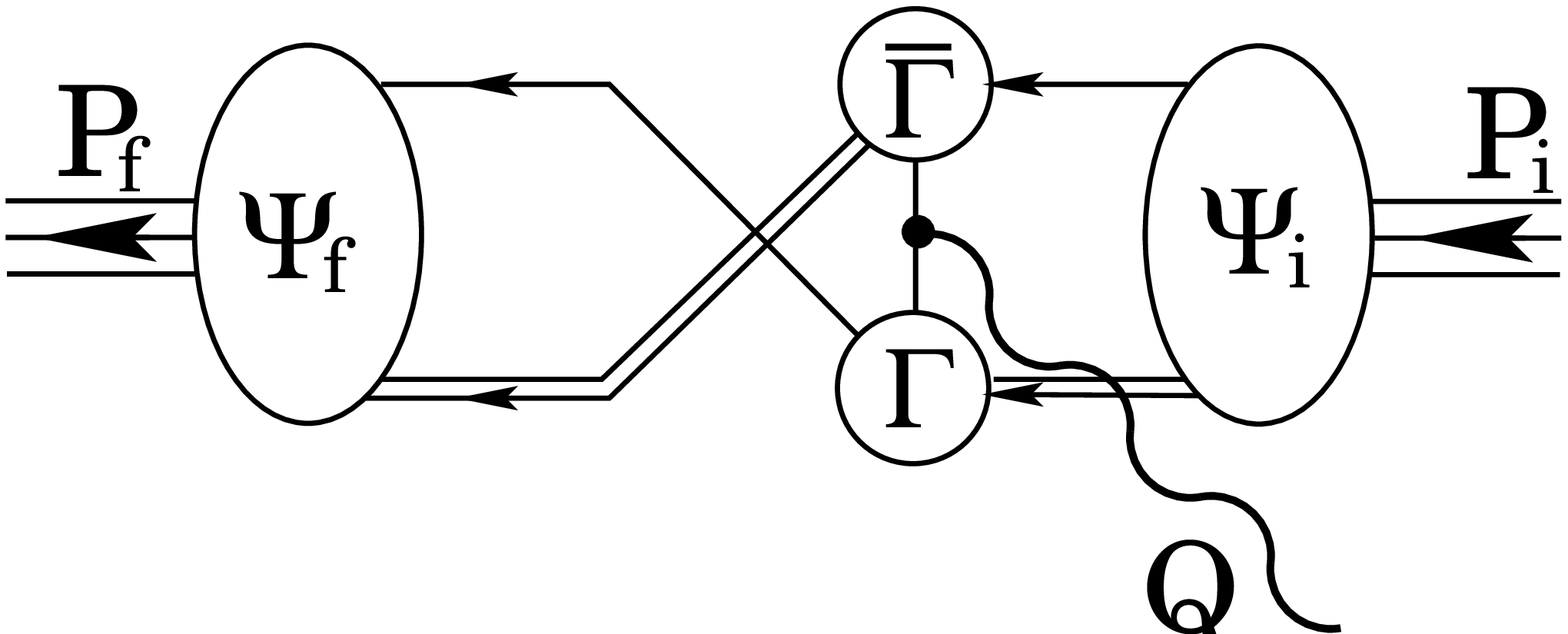}}
\end{minipage}
\begin{minipage}[t]{0.45\textwidth}
\rightline{\includegraphics[width=0.90\textwidth]{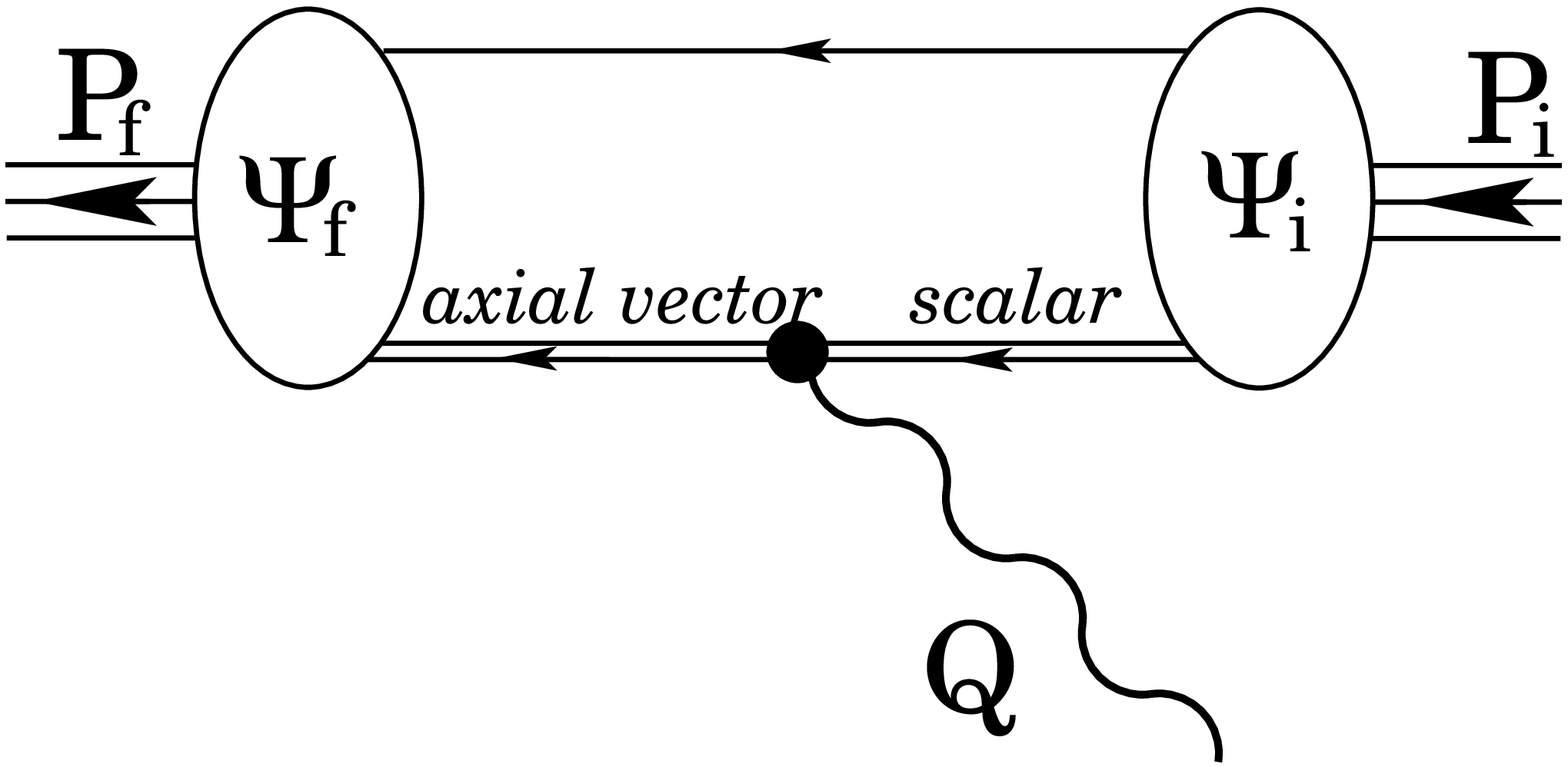}}
\end{minipage}\vspace*{3ex}

\begin{minipage}[t]{0.45\textwidth}
\leftline{\includegraphics[width=0.90\textwidth]{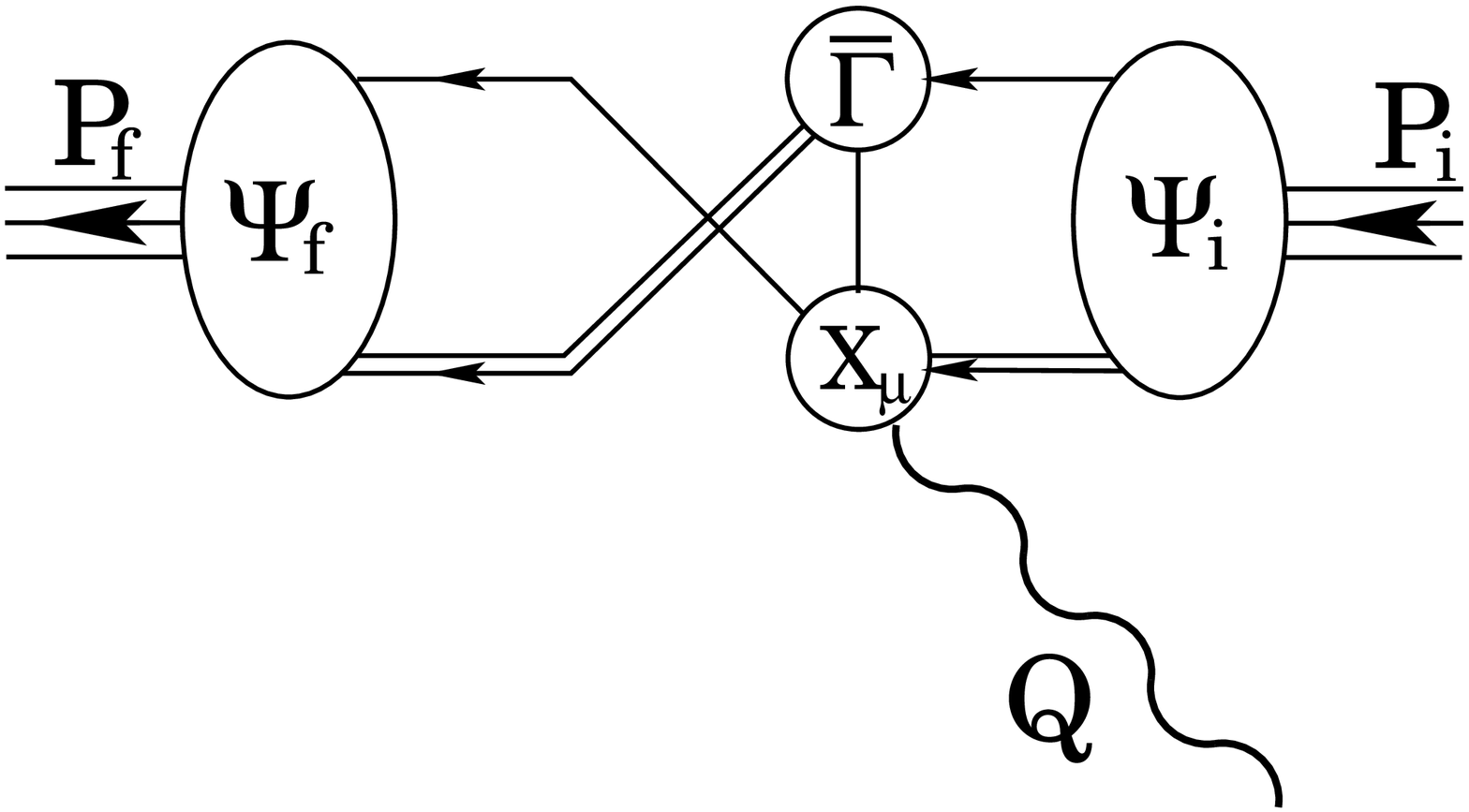}}
\end{minipage}
\begin{minipage}[t]{0.45\textwidth}
\rightline{\includegraphics[width=0.90\textwidth]{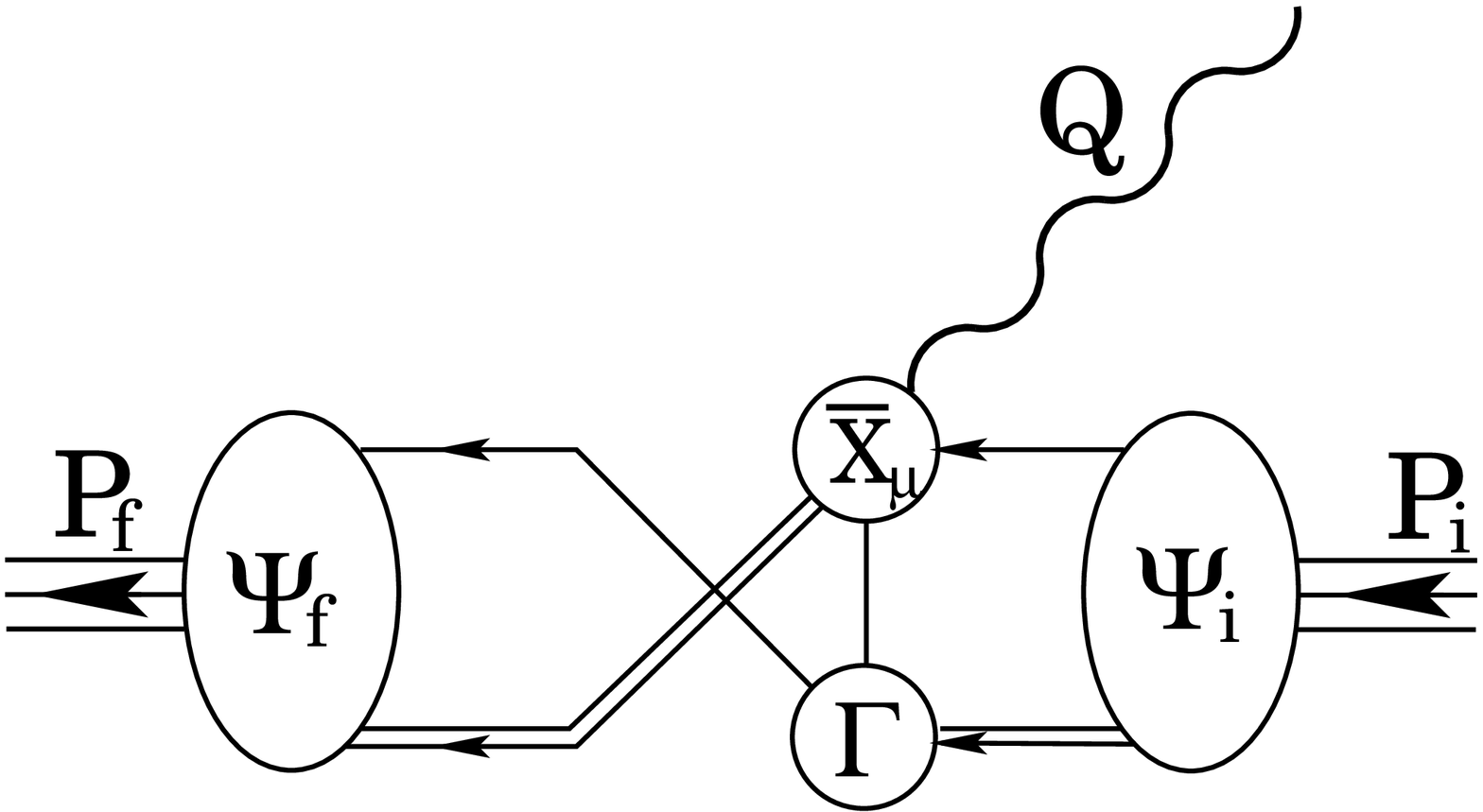}}
\end{minipage}
\end{minipage}
\caption{\label{vertex} Nucleon-photon vertex which ensures a conserved current for on-shell nucleons described by the Faddeev amplitudes, $\Psi_{i,f}$, calculated in Sec.\,\protect\ref{Sec:Faddeev}.  The single line represents $S(p)$, the dressed-quark propagator, Sec.\,\protect\ref{subsubsec:S}, and the double line, the diquark propagator, Sec.\,\protect\ref{qqprop}; $\Gamma$ is the diquark Bethe-Salpeter amplitude, Sec.\,\protect\ref{qqBSA}; and the remaining vertices are described in Secs.\,\protect\ref{Diag1}--\protect\ref{X56}: the top-left image is diagram~1; the top-right, diagram~2; and so on,  with the bottom-right image, diagram~6.}
\end{figure}

In Eq.\,(\ref{Jnucleon}), $\Lambda_\mu$ is the nucleon-photon vertex, which we construct following the systematic procedure of Ref.\,\cite{oettelpichowsky}.  This approach has the merit of automatically providing a conserved current for on-shell nucleons described by Faddeev amplitudes of the type we have calculated.  Moreover, the canonical normalisation condition for the nucleons' Faddeev amplitude is equivalent to requiring $F_1(Q^2=0)=1$ for the proton.  The vertex has six terms, which are depicted in Fig.~\ref{vertex} and expressed explicitly in \ref{appB}~Nucleon-Photon Vertex.  Here we describe the key elements in the construction. 

\subsection{Diagram~1}
\label{Diag1}
This represents the photon coupling directly to the bystander quark and is obtained explicitly from Eqs.\,(\ref{ABcurrent}) \& (\ref{B1}).  It is a necessary condition for current conservation that the quark-photon vertex satisfy the Ward-Takahashi identity:
\begin{equation}
\label{vwti}
Q_\mu \, i\Gamma_\mu(\ell_1,\ell_2) = S^{-1}(\ell_1) - S^{-1}(\ell_2)\,,
\end{equation}
where $Q=\ell_1-\ell_2$ is the photon momentum flowing into the vertex.  Since the quark is dressed, Sec.\,\ref{subsubsec:S}, the vertex is not bare; i.e., $\Gamma_\mu(\ell_1,\ell_2) \neq \gamma_\mu$.  It can be obtained by solving an inhomogeneous Bethe-Salpeter equation, which was the procedure adopted in the DSE calculation that successfully predicted the electromagnetic pion form factor \cite{pieterpion}.  However, since we have parametrised $S(p)$, we follow Ref.~\cite{blochff} and write \cite{bc80}
\begin{equation}
\label{bcvtx}
i\Gamma_\mu(\ell_1,\ell_2)  =  
i\Sigma_A(\ell_1^2,\ell_2^2)\,\gamma_\mu +
2 k_\mu \left[i\gamma\cdot k_\mu \,
\Delta_A(\ell_1^2,\ell_2^2) + \Delta_B(\ell_1^2,\ell_2^2)\right] \!;
\end{equation}
with $k= (\ell_1+\ell_2)/2$, $Q=(\ell_1-\ell_2)$ and
\begin{equation}
\Sigma_F(\ell_1^2,\ell_2^2) = \sfrac{1}{2}\,[F(\ell_1^2)+F(\ell_2^2)]\,,\;
\Delta_F(\ell_1^2,\ell_2^2) =
\frac{F(\ell_1^2)-F(\ell_2^2)}{\ell_1^2-\ell_2^2}\,,
\label{DeltaF}
\end{equation}
where $F= A, B$; viz., the scalar functions in Eq.\,(\ref{SpAB}).  It is
critical that $\Gamma_\mu$ in Eq.\ (\ref{bcvtx}) satisfies Eq.\ (\ref{vwti})
and very useful that it is completely determined by the dressed-quark
propagator.  Following Ref.\,\cite{cdrpion}, this \textit{Ansatz} has been used fruitfully in many hadronic applications.  Its primary defect is the omission of pion cloud contributions; e.g., Ref.\,\cite{bender}, but since one of our goals is to draw attention to consequences of that omission, this fault is herein a virtue.

\subsection{Diagram~2} 
\label{dqff}
This figure depicts the photon coupling directly to a diquark correlation, for which the explicit expression is obtained from Eqs.\,(\ref{ABcurrent}) \&  (\ref{B2}).  In the case of a scalar correlation, the general form of the diquark-photon vertex is
\begin{equation}
\Gamma_\mu^{0^+}(\ell_1,\ell_2) = 2\, k_\mu\, f_+(k^2,k\cdot Q,Q^2) + Q_\mu  \, f_-(k^2,k\cdot Q,Q^2)\,,
\end{equation}
and it must satisfy the Ward-Takahashi identity: 
\begin{equation}
\label{VWTI0}
Q_\mu \,\Gamma_\mu^{0^+}(\ell_1,\ell_2) = \Pi^{0^+}(\ell_1^2)  - \Pi^{0^+}(\ell_2^2)\,,\; \Pi^{J^P}(\ell^2) = \{\Delta^{J^P}(\ell^2)\}^{-1}.
\end{equation} 
The evaluation of scalar diquark elastic electromagnetic form factors in Ref.\,\cite{marisdqff} is a first step toward calculating this vertex.  However, in providing only an on-shell component, it is insufficient for our requirements.  We therefore adapt Eq.\,(\ref{bcvtx}) to this case and write
\begin{equation}
\label{Gamma0plus}
\Gamma_\mu^{0^+}(\ell_1,\ell_2) =  k_\mu\,
\Delta_{\Pi^{0^+}}(\ell_1^2,\ell_2^2)\,,
%
%\frac{k_\mu}{ k\cdot Q}\left\{ \left[\Delta^{0^+}(\ell_1^2)\right]^{-1} - \left[\Delta^{0^+}(\ell_2^2)\right]^{-1}\right\},
\end{equation}  
which is the minimal \textit{Ansatz} that: satisfies Eq.\,(\ref{VWTI0}); is completely determined by quantities introduced already; and is free of kinematic singularities.  It implements $f_- \equiv 0$, which is a requirement for elastic form factors, and guarantees a valid normalisation of electric charge; viz., 
\begin{equation}
\lim_{\ell^\prime\to \ell} \Gamma_\mu^{0^+}(\ell^\prime,\ell) = 2 \, \ell_{\mu} \, \frac{d}{d\ell^2}\, \Pi^{0^+}(\ell^2) \stackrel{\ell^2\sim 0}{=} 2 \, \ell_{\mu}\,,
\end{equation}
owing to Eq.\,(\ref{DQPropConstr}).  NB.\ We have factored the fractional diquark charge, which therefore appears subsequently in our calculations as a simple multiplicative factor. 

For the case in which the struck diquark correlation is axial-vector and the scattering is elastic, the vertex assumes the form \cite{HawesPichowsky99}:\,\footnote{If the scattering is inelastic the general form of the vertex involves eight scalar functions \protect\cite{Salam64}.  Absent further constraints and input, we ignore the additional structure in this \textit{Ansatz}.}
\begin{equation}
\label{AXDQGam}
\Gamma^{1^+}_{\mu\alpha\beta}(\ell_1,\ell_2) 
= -\sum_{i=1}^{3} \Gamma^{\rm [i]}_{\mu\alpha\beta}(\ell_1,\ell_2)\,,
\end{equation}
with ($T_{\alpha\beta}(\ell) = \delta_{\alpha\beta} - \ell_\alpha \ell_\beta/\ell^2$)
\begin{eqnarray}
\label{AXDQGam1}
\Gamma^{\rm [1]}_{\mu\alpha\beta}(\ell_1,\ell_2) 
&=& (\ell_1+\ell_2)_\mu \, T_{\alpha\lambda}(\ell_1) \, T_{\lambda\beta}(\ell_2)\; F_1(\ell_1^2,\ell_2^2)\,,
\\
\label{AXDQGam2}
\Gamma^{\rm [2]}_{\mu\alpha\beta}(\ell_1,\ell_2)
&=& \left[ T_{\mu\alpha}(\ell_1)\, T_{\beta\rho}(\ell_2) \, \ell_{1 \rho}
+ T_{\mu\beta}(\ell_2) \, T_{\alpha\rho}(\ell_1) \, \ell_{2\rho} \right] F_{2}(\ell_1^2,\ell_2^2) \,,
\\ 
\label{AXDQGam3}
\Gamma^{\rm [3]}_{\mu\alpha\beta}(\ell_1,\ell_2)
&=& -\frac{1}{2 m_{1^+}^2}\, (\ell_1+\ell_2)_\mu\, T_{\alpha\rho}(\ell_1)\, \ell_{2 \rho}
\, T_{\beta\lambda}(\ell_2)\, \ell_{1 \lambda}\; F_{3}(\ell_1^2,\ell_2^2) \,.
\end{eqnarray}
This vertex satisfies:
\begin{equation}
\ell_{1\alpha} \, \Gamma^{1^+}_{\mu\alpha\beta}(\ell_1,\ell_2) = 0 = 
\Gamma^{1^+}_{\mu\alpha\beta}(\ell_1,\ell_2) \, \ell_{2\beta} \,,
\end{equation}
which is a general requirement of the elastic electromagnetic vertex of axial-vector bound states and guarantees that the interaction does not induce a pseudoscalar component in the axial-vector correlation.  We note that the electric, magnetic and quadrupole form factors of an axial-vector bound state are expressed \cite{HawesPichowsky99}
\begin{eqnarray}
\label{GEDQ}
& &
G_{\cal E}^{1^+}(Q^2) = F_1 + \sfrac{2}{3}\, \tau_{1^+}\, 
G_{\cal Q}^{1^+}(Q^2) \,, \; \tau_{1^+} = \frac{Q^2}{ 4 \,m_{1^{+}}^{2}}
\\
\label{GMDQ}
& &
G_{\cal M}^{1^+}(Q^2) = - F_2(Q^2) ~,
\\
& &
\label{GQDQ}
G_{\cal Q}^{1^+}(Q^2) = F_1(Q^2) + F_2(Q^2) + \left( 1 + \tau_{1^+}\right) F_3(Q^2) \,.
\end{eqnarray}

Extant knowledge of the form factors in Eqs.\,(\ref{AXDQGam})--(\ref{AXDQGam3}) is limited and thus one has little information about even this rudimentary vertex model.  Hence, we employ the following \textit{Ans\"atze}:
\begin{eqnarray}
\label{AnsatzF1}
F_{1}(\ell_1^2,\ell_2^2) &=& \Delta_{\Pi^{1^+}}(\ell_1^2,\ell_2^2)\,, \\
\label{AnsatzF2}
F_{2}(\ell_1^2,\ell_2^2) &=& -\, F_{1} + 
(1-\tau_{1^+}) \,( \tau_{1^+} F_{1}+1 - \mu_{1^+})\, d(\tau_{1^+}) \\
\label{AnsatzF3}
F_{3}(\ell_1^2,\ell_2^2) &=& -\,(\chi_{1^+}\,(1- \tau_{1^+})\,d(\tau_{1^+})+F_1 + F_2)\, d(\tau_{1^+})\,,
\end{eqnarray}
with $d(x)=1/(1+x)^2$.  This construction ensures a valid electric charge normalisation for the axial-vector correlation; viz., 
\begin{equation}
\lim_{\ell^\prime \to\ell} \, \Gamma^{1^+}_{\mu\alpha\beta}(\ell^\prime,\ell) = T_{\alpha\beta}(\ell) \,\frac{d}{d\ell^2}\, \Pi^{1^+}(\ell^2) 
\stackrel{\ell^2\sim 0}{=}  T_{\alpha\beta}(\ell) \,2 \,\ell_{ \mu}\,,
\end{equation}
owing to Eq.\,(\ref{DQPropConstr}), and current conservation 
\begin{equation}
\lim_{\ell_2\to\ell_1} \, Q_\mu \Gamma^{1^+}_{\mu\alpha\beta}(\ell_1,\ell_2) = 0\,.
\end{equation}
The diquark's static electromagnetic properties follow: 
\begin{equation}
\label{pointp}
G_{\cal E}^{1^+}(0) = 1\,,\;
G_{\cal M}^{1^+}(0) = \mu_{1^+}\,,\;
G_{\cal Q}^{1^+}(0) = -\chi_{1^+}\,.
\end{equation}
For a pointlike axial-vector: $\mu_{1^+}=2$; and $\chi_{1^+}=1$, which corresponds to an oblate charge distribution.  In addition, Eqs.\,(\ref{AXDQGam})--(\ref{AXDQGam3}) with Eqs.\,(\ref{AnsatzF1})--(\ref{AnsatzF3}) realise the constraints of Ref.\,\cite{brodskyhiller92}; namely, independent of the values of $\mu_{1^+}$ \& $\chi_{1^+}$, the form factors assume the ratios
\begin{equation}
\label{pQCDavdq}
G_{\cal E}^{1^+}(Q^2): G_{\cal M}^{1^+}(Q^2): G_{\cal Q}^{1^+}(Q^2)
\stackrel{Q^2\to \infty}{=} (1 - \sfrac{2}{3} \tau_{1^+}) : 2 : - 1 \,.
\end{equation}

\subsection{Diagram~3}
This image depicts a photon coupling to the quark that is exchanged as one diquark breaks up and another is formed.  While this is the first two-loop diagram we have described, no new elements appear in its specification: the dressed-quark-photon vertex was discussed in Sec.~\ref{Diag1}.  The explicit expression for this diagram's contribution to the nucleons' form factors is obtained from Eqs.\,(\ref{ABcurrent}) \& (\ref{B3}).  

It is noteworthy that the process of quark exchange provides the attraction necessary in the Faddeev equation to bind the nucleon.  It also guarantees that the Faddeev amplitude has the correct antisymmetry under the exchange of any two dressed-quarks.  This key feature is absent in models with elementary (noncomposite) diquarks.

\subsection{Diagram~4}
\label{kTsec}
This differs from Diagram~2 in expressing the contribution to the nucleons' form factors owing to an electromagnetically induced transition between scalar and axial-vector diquarks: the explicit expression is described in connection with Eq.\,(\ref{B2}).  The transition vertex is a rank-2 pseudotensor, kindred to the matrix element describing the $\rho\, \gamma^\ast \pi^0$  transition \cite{maristransition}, and can therefore be expressed 
\begin{equation}
\label{SAPhotVertex}
\Gamma_{SA}^{\gamma\alpha}(\ell_1,\ell_2) = -\Gamma_{AS}^{\gamma\alpha}(\ell_1,\ell_2) 
= \frac{i}{M_N} \, {\cal T}(\ell_1,\ell_2) \, \varepsilon_{\gamma\alpha\rho\lambda}\ell_{1\rho} \ell_{2 \lambda}\,,
\end{equation}
where $\gamma$, $\alpha$ are, respectively, the vector indices of the photon and axial-vector diquark.  For simplicity we proceed under the assumption that
\begin{equation}
{\cal T}(\ell_1,\ell_2) = \kappa_{\cal T}\,;
\end{equation}
viz., a constant, for which a typical value is \cite{oettelff}: 
\begin{equation}
\label{kTbest}
\kappa_{\cal T} \sim 2\,.
\end{equation}

In the nucleons' rest frame, a conspicuous piece of the Faddeev amplitude that describes an axial-vector diquark inside the bound state can be characterised as containing a bystander quark whose spin is antiparallel to that of the nucleon, with the axial-vector diquark's parallel.  The interaction pictured in this diagram does not affect the bystander quark but the transformation of an axial-vector diquark into a scalar effects a flip of the quark spin within the correlation.  After this transformation, the spin of the nucleon must be formed by summing the spin of the bystander quark, which is still aligned antiparallel to that of the nucleon, and the orbital angular momentum between that quark and the scalar diquark.\footnote{A less prominent component of the amplitude has the bystander quark's spin parallel to that of the nucleon while the axial-vector diquark's is antiparallel: this $q^\uparrow \oplus (qq)_{1^+}^{\downarrow} $ system has one unit of angular momentum.  That momentum is absent in the $q^\uparrow \oplus (qq)_{0^+}$ system.  Other combinations also contribute via Diagram~3 but all mediated processes inevitably require a modification of spin and/or angular momentum.}   This argument, while not sophisticated, does motivate an expectation that Diagram~4 will strongly impact on the nucleons' magnetic form factors.

\subsection{Diagrams~5 \& 6}
\label{X56}
These two-loop diagrams are the so-called ``seagull'' terms, which appear as partners to Diagram~3 and arise because binding in the nucleons' Faddeev equations is effected by the exchange of \textit{nonpointlike} diquark correlations \cite{oettelpichowsky}.  The explicit expression for their contribution to the nucleons' form factors is obtained from Eqs.\,(\ref{ABcurrent}) and (\ref{B5}).  The new elements in these diagrams are the couplings of a photon to two dressed-quarks as they either separate from (Diagram~5) or combine to form (Diagram~6) a diquark correlation.  As such they are components of the five point Schwinger function which describes the coupling of a photon to the quark-quark scattering kernel.  This Schwinger function could be calculated, as is evident from the recent computation of analogous Schwinger functions relevant to meson observables \cite{marisgppp}.  However, such a calculation provides valid input only when a uniform truncation of the DSEs has been employed to calculate each of the elements described hitherto.  We must instead employ an algebraic parametrisation \cite{oettelpichowsky}, which for Diagram~5 reads
\begin{eqnarray}
\nonumber
X^{J^P}_\mu(k,Q) & =&  e_{\rm by}\,\frac{4 k_\mu- Q_\mu}{4 k\cdot Q - Q^2}\,\left[\Gamma^{J^P}\!(k-Q/2)-\Gamma^{J^P}\!(k)\right]\\
& +& e_{\rm ex}\,\frac{4 k_\mu+ Q_\mu}{4 k\cdot Q + Q^2}\,\left[\Gamma^{J^P}\!(k+Q/2)-\Gamma^{J^P}\!(k)\right], \label{X5}
\end{eqnarray}
with $k$ the relative momentum between the quarks in the initial diquark, $e_{\rm by}$ the electric charge of the quark which becomes the bystander and $e_{\rm ex}$, the charge of the quark that is reabsorbed into the final diquark.  Diagram~6 has
\begin{eqnarray}
\nonumber
\bar X^{J^P}_\mu(k,Q) & =&  -\,e_{\rm by}\,\frac{4 k_\mu- Q_\mu}{4 k\cdot Q - Q^2}\,\left[\bar\Gamma^{J^P}\!(k+Q/2)-\bar\Gamma^{J^P}\!(k)\right]\\
& -& e_{\rm ex}\,\frac{4 k_\mu+ Q_\mu}{4 k\cdot Q + Q^2}\,\left[\bar\Gamma^{J^P}\!(k-Q/2)-\bar\Gamma^{J^P}\!(k)\right], \label{X6}
\end{eqnarray}
where $\bar\Gamma^{J^P}\!(\ell)$ is the charge-conjugated amplitude, Eq.\,(\ref{chargec}).  Plainly, these terms vanish if the diquark correlation is represented by a momentum-independent Bethe-Salpeter-like amplitude; i.e., the diquark is pointlike.

It is naturally possible to use more complicated \textit{Ans\"atze}.  However, like Eq.\,(\ref{Gamma0plus}), Eqs.\,(\ref{X5}) \& (\ref{X6}) are simple forms, free of kinematic singularities and sufficient to ensure the nucleon-photon vertex satisfies the Ward-Takahashi identity when the composite nucleon is obtained from the Faddeev equation.

\section{Nucleon Electromagnetic Form Factors}
\label{FFs}
%...commentary
\subsection{Exegesis}
In order to place the calculation of baryon observables on the same footing as the study of mesons, the proficiency evident in Refs.\,\cite{marisrev,marisgppp} will need to be applied to every line and vertex that appears in Fig.\,\ref{vertex}.  While that is feasible, it requires a considerable investment of resources.  In the meantime, herein we present a study whose merits include a capacity to: explore the potential of the Faddeev equation truncation of the baryon three-body problem; and elucidate the role of additional correlations, such as those associated with pseudoscalar mesons.

It is worthwhile to epitomise our input before presenting the results.  One element is the dressed-quark propagator, Sec.\,\ref{subsubsec:S}.  The form we use \cite{mark} both anticipated and expresses the features that are now known to be true \cite{bhagwat,maris}.  It carries no free parameters, because its behaviour was fixed in analyses of meson observables, and is basic to a description of light- and heavy-quark mesons that is accurate to better than 10\% \cite{mishasvy}.

We proposed that the nucleon is at heart composed of a dressed-quark and nonpointlike diquark with binding effected by an iterated exchange of roles between the bystander and diquark-participant quarks.  This picture is realised via a Poincar\'e covariant Faddeev equation, Sec.\,\ref{ANDelta}, which incorporates scalar and axial-vector diquark correlations.  There are two parameters, Secs.\,\ref{qqBSA} \& \ref{qqprop}: the mass-scales associated with these correlations.  They are fixed by fitting to specified nucleon and $\Delta$ masses, Sec.\,\ref{NDmasses}, and thus at this point there are still no free parameters with which to influence the nucleons' form factors.

With the constituents and the bound states' structure defined, only a specification of the nucleons' electromagnetic interaction remained.  Its formulation was guided almost exclusively by a requirement that the nucleon-photon vertex satisfy a Ward-Takahashi identity.  Since the scalar diquark's electromagnetic properties are readily resolved, our result, Fig.\,\ref{vertex}, depends on three parameters that are all tied to properties of the axial-vector diquark correlation: $\mu_{1^+}$ \& $\chi_{1^+}$, respectively, the axial-vector diquarks' magnetic dipole and electric quadrupole moments; and $\kappa_{\cal T}$, the strength of electromagnetic axial-vector $\leftrightarrow$ scalar diquark transitions.  Hence, with our calculations we exhibit and interpret the dependence of the nucleons' form factors on these three parameters, and also on the nucleons' intrinsic quark structure as expressed in the Poincar\'e covariant Faddeev amplitudes. 

\subsection{Calculated Results and Discussion}
\label{results}
\subsubsection{Static properties and form factors}
The nucleons' charge and magnetic radii are defined respectively as
\begin{equation}
r_{N}^2:= \left.- \, 6\,\frac{d}{ds} \ln G_E^{N}(s) \right|_{s=0},\;
(r_{N}^\mu)^2:= \left.- \, 6\,\frac{d}{ds} \ln G_M^{N}(s) \right|_{s=0},
\end{equation}
where $N=n,p$; and in Table~\ref{radii} we report charge radii calculated for a range of values of the parameters that characterise the axial-vector diquarks' electromagnetic form factors, Sec.\,\ref{dqff}, centred on the point-particle values of $\mu_{1^+}=2$ \& $\chi_{1^+}=1$, Eq.\,(\protect\ref{pointp}), and $\kappa_{\cal T} = 2$, Eq.\,(\ref{kTbest}).  The radii, particularly that of the neutron, are most sensitive to changes in the axial-vector diquarks' electric quadrupole moment, $\chi_{1^+}$.  This is not surprising given that $\chi_{1^+}$ is the only model parameter that speaks directly of the axial-vector diquarks' electric charge distribution.  The radii's insensitivity to $\kappa_{\cal T}$, the strength of the scalar $\leftrightarrow$ axial-vector transition, is concordant with the discussion in Sec.\,\ref{kTsec}.  With the reference values given in Eqs.\,(\ref{pointp}) \& (\ref{kTbest}), Set~A underestimates the proton radius by 30\% and the magnitude of the neutron radius by 43\%, while for Set~B these differences are 32\% and 50\%, respectively.  

\begin{table}[t]
\begin{center}
\caption{\label{radii} Charge radii, in fm, calculated using the diquark mass-scale parameters in Table~\protect\ref{ParaFix} for a range of axial-vector-diquark--photon vertex parameters, centred on the point-particle values of $\mu_{1^+}=2$ \& $\chi_{1^+}=1$, Eq.\,(\protect\ref{pointp}), and $\kappa_{\cal T} = 2$, Eq.\,(\ref{kTbest}).  Columns labelled $\sigma$ give the percentage-difference from results obtained with the reference values.  $r_n:= -\sqrt{-\langle r_n^2\rangle}$.  Values inferred from experiment are \cite{drechsel}: $r_p = 0.847$ \& $r_n = - 0.336$.}
\begin{tabular*}{1.0\textwidth}{c@{\extracolsep{0ptplus1fil}}
c@{\extracolsep{0ptplus1fil}}
c@{\extracolsep{0ptplus1fil}}|c@{\extracolsep{0ptplus1fil}} 
c@{\extracolsep{0ptplus1fil}}c@{\extracolsep{0ptplus1fil}}
c@{\extracolsep{0ptplus1fil}}c@{\extracolsep{0ptplus1fil}}
|c@{\extracolsep{0ptplus1fil}}
c@{\extracolsep{0ptplus1fil}}c@{\extracolsep{0ptplus1fil}}
c@{\extracolsep{0ptplus1fil}}}
 \multicolumn{3}{c}{} &\multicolumn{5}{c}{Set~A} & \multicolumn{4}{c}{Set~B} \\\hline
$\mu_{1^+}$ & $\chi_{1^+}$ & $\kappa_{\cal T}$ & 
$r_p$ & \rule{0em}{2.5ex}$\sigma_{r_p}^A$  & $r_n$ & $\sigma_{r_n}^A$ & & $r_p$ & $\sigma_{r_p}^B$ & $r_n$ & $\sigma_{r_n}^B$ \\\hline
%
%1 & 1 & 0 & 0.596 & -1.3  & 0.178 & -4.8 && 0.595 & ~0.3 & 0.167 & ~~1.2 \\
%2 & 1 & 0 & 0.604 & ~~~   & 0.187 & ~~~  && 0.593 & ~~~  & 0.165 & ~~~ \\
%3 & 1 & 0 & 0.611 & ~1.2  & 0.195 & ~4.3 && 0.592 & -0.2 & 0.162 & ~-1.8 \\\hline
%
%2 & 0 & 0 & 0.591 & -2.2  & 0.172 & -8.0 && 0.573 & -3.4 & 0.138 & -16.4 \\
%2 & 2 & 0 & 0.617 & ~2.2  & 0.202 & ~8.0 && 0.613 & ~3.4 & 0.186 & ~12.7 \\\hline
%
1 & 1 & 2 & 0.599 & -1.2  & 0.185 & -4.1 && 0.596 & ~0.2 & 0.171 & ~~1.2 \\
2 & 1 & 2 & 0.606 & ~~~~  & 0.193 & ~~~~ && 0.595 & ~~~~ & 0.169 & ~~~~ \\
3 & 1 & 2 & 0.614 & ~1.3  & 0.200 & ~3.6 && 0.593 & -0.3 & 0.167 & ~-1.2 \\\hline
2 & 0 & 2 & 0.593 & -2.2  & 0.179 & -7.3 && 0.575 & -3.4 & 0.145 & -14.2 \\
%2 & 1 & 2 & 0.606 & ~~~~  & 0.193 & ~3.2 && 0.595 & ~~~~ & 0.169 & ~~3.0 \\
2 & 2 & 2 & 0.620 & ~2.3  & 0.205 & ~6.2 && 0.614 & ~3.2 & 0.191 & ~13.0 \\\hline
2 & 1 & 1 & 0.606 & ~0.0  & 0.189 & -2.1 && 0.595 & ~0.0 & 0.167 & ~-1.2 \\
%2 & 1 & 2 & 0.606 & ~~~~  & 0.193 & ~3.2 && 0.595 & ~~~~ & 0.169 & ~~3.0 \\
2 & 1 & 3 & 0.606 & ~0.0  & 0.196 & ~1.6 && 0.595 & ~0.0 & 0.172 & ~~1.8 \\\hline
\end{tabular*}
\end{center}
\end{table}

\begin{table}[tb]
\begin{center}
\caption{\label{magradii} Magnetic radii, in fm, calculated with the diquark mass-scales in Table~\protect\ref{ParaFix} and the parameter range described in Table~\protect\ref{radii}.
%using the diquark mass-scale parameters in Table~\protect\ref{ParaFix} for a range of axial-vector-diquark--photon vertex parameters, centred on the point-particle values for $\mu_{1^+}$ \& $\chi_{1^+}$, Eq.\,(\protect\ref{pointp}), and $\kappa_{\cal T} = 2$, Eq.\,(\ref{kTbest}).  
Columns labelled $\sigma$ give the percentage-difference from results obtained with the reference values: $\mu_{1^+}=2$, $\chi_{1^+}=1$,  $\kappa_{\cal T} = 2$.  Values inferred from experiment are \cite{drechsel}: $r_p^{\mu} = 0.836$ \& $r_n^\mu=0.889$.}
\begin{tabular*}{1.0\textwidth}{c@{\extracolsep{0ptplus1fil}}c@{\extracolsep{0ptplus1fil}}
c@{\extracolsep{0ptplus1fil}}|c@{\extracolsep{0ptplus1fil}} 
c@{\extracolsep{0ptplus1fil}}c@{\extracolsep{0ptplus1fil}}
c@{\extracolsep{0ptplus1fil}}c@{\extracolsep{0ptplus1fil}}|c@{\extracolsep{0ptplus1fil}}
c@{\extracolsep{0ptplus1fil}}c@{\extracolsep{0ptplus1fil}}c@{\extracolsep{0ptplus1fil}}}
 \multicolumn{3}{c}{} & \multicolumn{5}{c}{Set~A} & \multicolumn{4}{c}{Set~B} \\\hline
$\mu_{1^+}$ & $\chi_{1^+}$ & $\kappa_{\cal T}$ & $r_p^\mu$ & \rule{0em}{2.5ex}$\sigma_{r_p^\mu}^A$ & $r_n^\mu$ & $\sigma_{r_n^\mu}^A$ & & $r_p^\mu$ & $\sigma_{r_p^\mu}^B$ & $r_n^\mu$ & $\sigma_{r_n^\mu}^B$ \\\hline
%
%1 & 1 & 0 & 0.596 & -1.3  & 0.178 & -4.8 && 0.595 & ~0.3 & 0.167 & ~~1.2 \\
%2 & 1 & 0 & 0.604 & ~~~   & 0.187 & ~~~  && 0.593 & ~~~  & 0.165 & ~~~ \\
%3 & 1 & 0 & 0.611 & ~1.2  & 0.195 & ~4.3 && 0.592 & -0.2 & 0.162 & ~-1.8 \\\hline
%
%2 & 0 & 0 & 0.591 & -2.2  & 0.172 & -8.0 && 0.573 & -3.4 & 0.138 & -16.4 \\
%2 & 2 & 0 & 0.617 & ~2.2  & 0.202 & ~8.0 && 0.613 & ~3.4 & 0.186 & ~12.7 \\\hline
%
1 & 1 & 2 & 0.456 & -2.4  & 0.467 & -1.3 && 0.442 & -1.6 & 0.446 & -0.7 \\
2 & 1 & 2 & 0.467 & ~~~~  & 0.473 & ~~~~ && 0.449 & ~~~~ & 0.449 & ~~~~ \\
3 & 1 & 2 & 0.477 & ~2.1  & 0.478 & ~1.1 && 0.454 & ~1.1 & 0.454 & ~1.1 
\\\hline
2 & 0 & 2 & 0.467 & ~0.0  & 0.473 & ~0.0 && 0.449 & ~0.0 & 0.449 & ~0.0 \\
%2 & 1 & 2 & 0.467 & ~~~~  & 0.473 & ~~~~ && 0.449 & ~~~~ & 0.449 & ~~~~ \\
2 & 2 & 2 & 0.467 & ~0.0  & 0.473 & ~0.0 && 0.449 & ~0.0 & 0.449 & ~0.0 
\\\hline
2 & 1 & 1 & 0.470 & ~0.6  & 0.480 & ~1.3 && 0.453 & ~0.9 & 0.459 & ~2.2 \\
%2 & 1 & 2 & 0.467 & ~~~~  & 0.473 & ~~~~ && 0.449 & ~~~~ & 0.449 & ~~~~ \\
2 & 1 & 3 & 0.465 & -0.4  & 0.472 & -0.2 && 0.445 & -0.9 & 0.446 & -0.7 
\\\hline
\end{tabular*}
\end{center}
\end{table}

Table~\ref{magradii} presents results for the nucleons' magnetic radii.  They are insensitive to the axial-vector diquarks' quadrupole moment but react to the diquarks' magnetic moment as one would anticipate: increasing in magnitude as $\mu_{1^+}$ increases.  Moreover, consistent with expectation, Sec.\,\ref{kTsec}, these radii also respond to changes in $\kappa_{\cal T}$, decreasing as this parameter is increased.  With the reference values in Eqs.\,(\ref{pointp}) \& (\ref{kTbest}), both Sets underestimate $r_N^\mu$ by approximately 40\%.

\begin{table}[bt]
\begin{center}
\caption{\label{moments} Magnetic moments, in nuclear magnetons, calculated with the diquark mass-scales in Table~\protect\ref{ParaFix} and the parameter range described in Table~\protect\ref{radii}.
%using the diquark mass-scale parameters in Table~\protect\ref{ParaFix} for a range of axial-vector-diquark--photon vertex parameters, centred on the point-particle values for $\mu_{1^+}$ \& $\chi_{1^+}$, Eq.\,(\protect\ref{pointp}), and $\kappa_{\cal T} = 2$, Eq.\,(\ref{kTbest}).  
Columns labelled $\sigma$ give the percentage-difference from results obtained with the reference values: $\mu_{1^+}=2$, $\chi_{1^+}=1$,  $\kappa_{\cal T} = 2$.  Experimental values are: $\kappa_p:=\mu_p-1 = 1.79$ \& $\mu_n = -1.91$.  }
\begin{tabular*}{1.0\textwidth}{c@{\extracolsep{0ptplus1fil}}
c@{\extracolsep{0ptplus1fil}}
c@{\extracolsep{0ptplus1fil}}|c@{\extracolsep{0ptplus1fil}} 
c@{\extracolsep{0ptplus1fil}}c@{\extracolsep{0ptplus1fil}}
c@{\extracolsep{0ptplus1fil}}c@{\extracolsep{0ptplus1fil}}
|c@{\extracolsep{0ptplus1fil}}
c@{\extracolsep{0ptplus1fil}}c@{\extracolsep{0ptplus1fil}}c@{\extracolsep{0ptplus1fil}}}
 \multicolumn{3}{c}{} & \multicolumn{5}{c}{Set~A} & \multicolumn{4}{c}{Set~B} \\\hline
$\mu_{1^+}$ & $\chi_{1^+}$ & $\kappa_{\cal T}$ & $\kappa_p$ & \rule{0em}{2.5ex}$\sigma_{\kappa_p}^A$ & $|\mu_n|$ & $\sigma_{|\mu_n|}^A$ && $\kappa_p$ & $\sigma_{\kappa_p}^B$ & $|\mu_n|$ & $\sigma_{|\mu_n|}^B$ \\\hline
%
%1 & 1 & 0 & 1.49 & -15.5  & 1.40 & ~-6.1 && 1.88 & -17.2 & 1.63 & ~-7.4 \\
%2 & 1 & 0 & 1.76 & ~~~    & 1.49 & ~~~  && 2.27 & ~~~   & 1.76 & ~~~ \\
%3 & 1 & 0 & 2.03 & ~15.7  & 1.58 & ~~6.1 && 2.66 & ~17.4 & 1.90 & ~~7.4 \\\hline
%
%2 & 0 & 0 & 1.76 & ~~0.0  & 1.49 & ~~0.0 && 2.27 & ~~0.0 & 1.76 & ~~0.0 \\
%2 & 2 & 0 & 1.76 & ~~0.0  & 1.49 & ~~0.0 && 2.27 & ~~0.0 & 1.76 & ~~0.0 \\\hline
%
1 & 1 & 2 & 1.79 & -15.3  & 1.70 & -5.1 && 2.24 & -21.9 & 2.00 & -6.2 \\
2 & 1 & 2 & 2.06 & ~~~~~  & 1.79 & ~~~~ && 2.63 & ~~~~~ & 2.13 & ~~~~ \\
3 & 1 & 2 & 2.33 & ~15.4  & 1.88 & ~5.1 && 3.02 & ~21.9 & 2.26 & ~6.1 \\\hline
2 & 0 & 2 & 2.06 & ~~0.0  & 1.79 & ~0.0 && 2.63 & ~~0.0 & 2.13 & ~0.0 \\
%2 & 1 & 2 & 2.06 & ~~~~~  & 1.79 & 20.0 && 2.63 & ~~~~~ & 2.13 & ~20.6 \\
2 & 2 & 2 & 2.06 & ~~0.0  & 1.79 & ~0.0 && 2.63 & ~~0.0 & 2.13 & ~0.0 \\\hline
2 & 1 & 1 & 1.91 & ~-8.4  & 1.64 & -8.4 && 2.45 & -10.1 & 1.95 & -8.5 \\
%2 & 1 & 2 & 2.06 & ~~~~~  & 1.79 & 20.0 && 2.63 & ~~~~~ & 2.13 & ~20.6 \\
2 & 1 & 3 & 2.21 & ~~8.4  & 1.85 & ~8.3 && 2.82 & ~10.1 & 2.31 & ~8.5 \\\hline
\end{tabular*}
\end{center}
\end{table}

Table~\ref{moments} lists results for the nucleons' magnetic moments.  They, too, are insensitive to the axial-vector diquarks' quadrupole moment but react to the diquarks' magnetic moment, increasing quickly in magnitude as $\mu_{1^+}$ increases.  As anticipated in Sec.\,\ref{kTsec}, the nucleons' moments respond strongly to alterations in the strength of the scalar $\leftrightarrow$ axial-vector transition, increasing rapidly as $\kappa_{\cal T}$ is increased.   Set~A, which is fitted to the experimental values of $M_N$ \& $M_\Delta$, describes the nucleons' moments quite well: $\kappa_p$ is 15\% too large; and $|\mu_n|$, 16\% too small.  On the other hand, Set~B, which is fitted to baryon masses that are inflated so as to make room for pion cloud effects, overestimates $\kappa_p$ by 47\% and $|\mu_n|$ by 18\%.

\begin{figure}[t]
\begin{minipage}{0.45\textwidth}
\centerline{\hspace*{2.5em}%
\includegraphics[width=0.95\textwidth,angle=270]{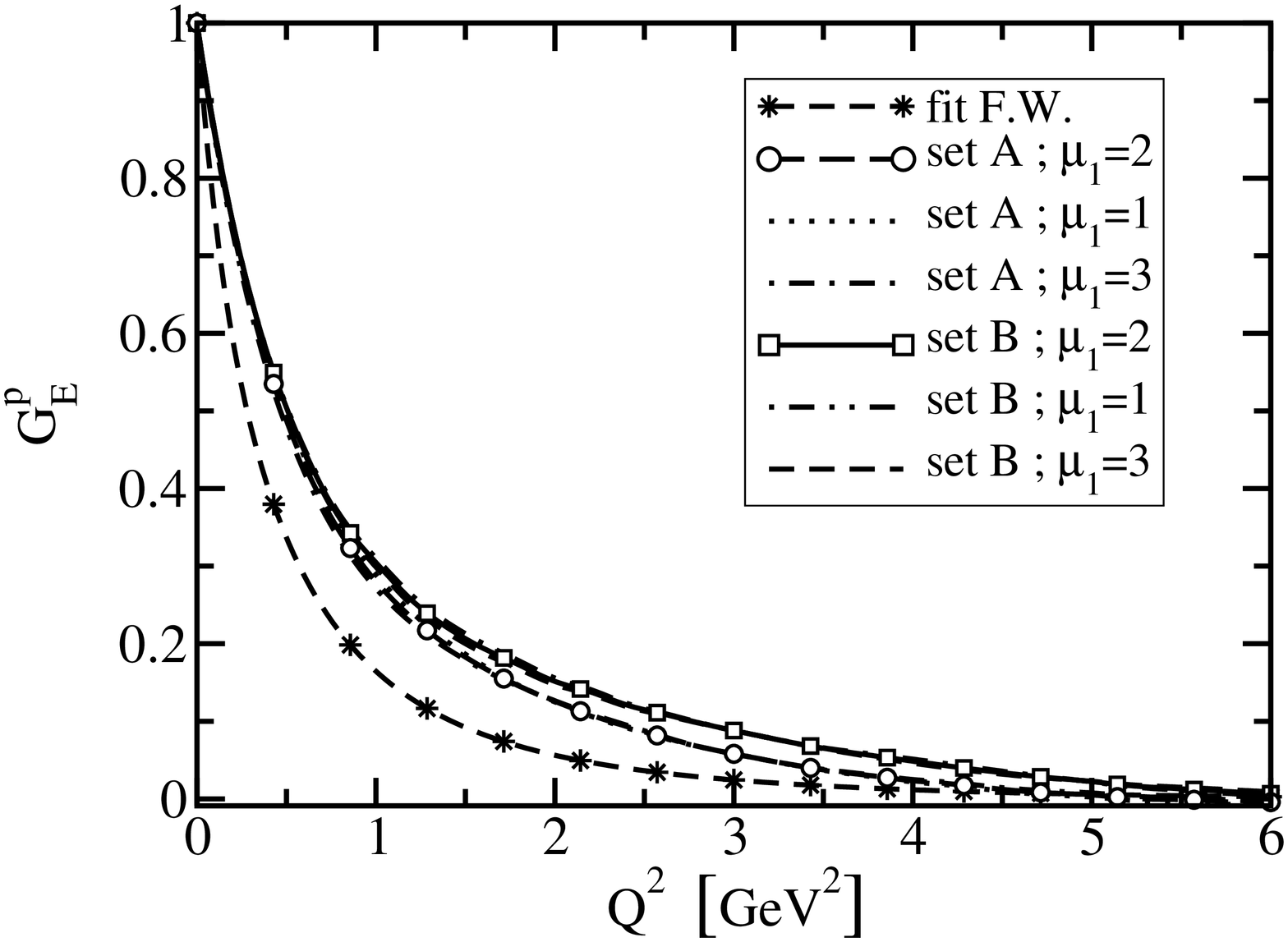}}
\end{minipage}
\hfill
\begin{minipage}{0.45\textwidth}
\centerline{\includegraphics[width=0.95\textwidth,angle=270]{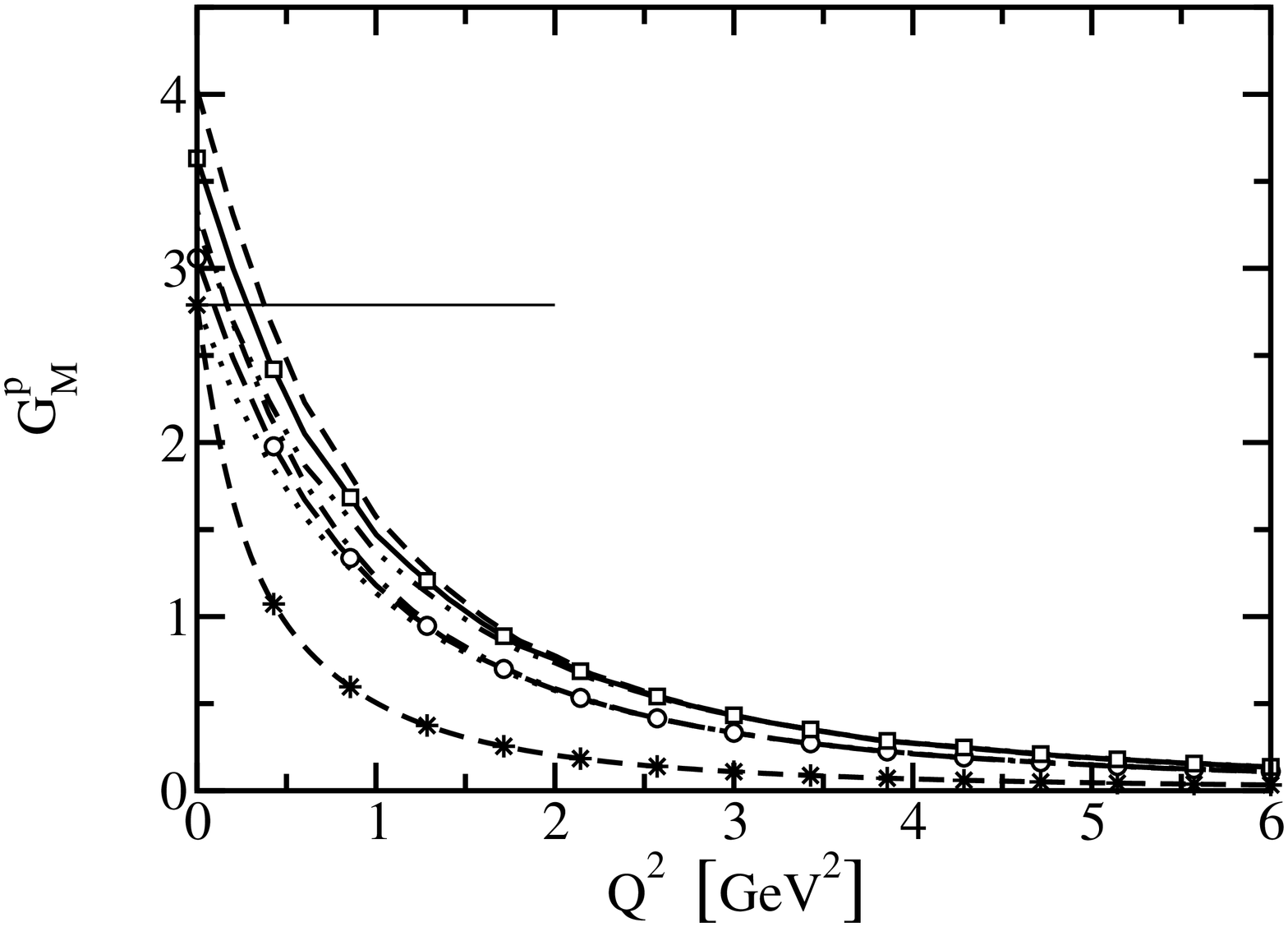}%
\hspace{1em}}
\end{minipage}
\\
\begin{minipage}{0.45\textwidth}
\centerline{\hspace*{2.5em}%
\includegraphics[width=0.95\textwidth,angle=270]{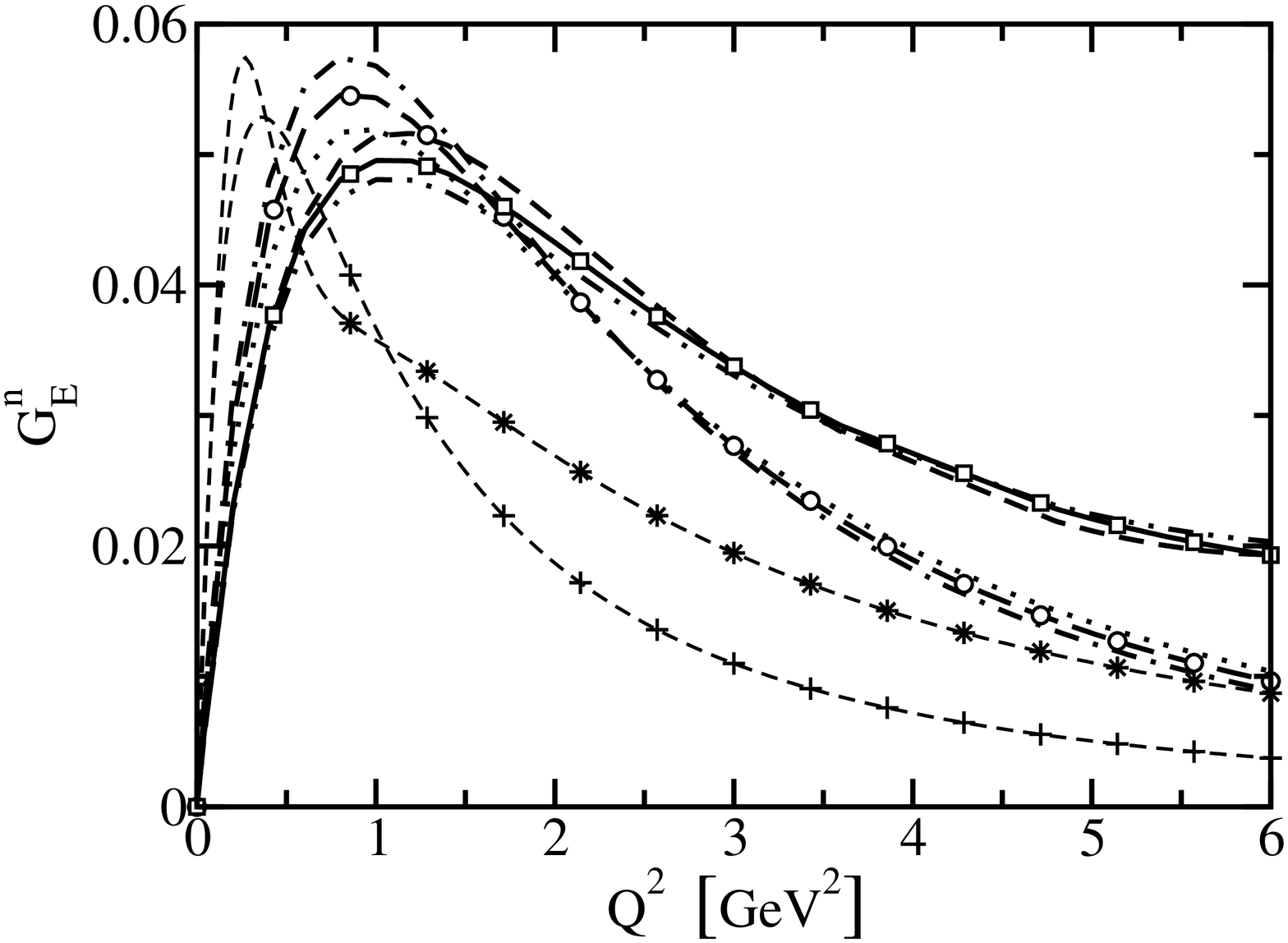}}
\end{minipage}
\hfill
\begin{minipage}{0.45\textwidth}
\centerline{\includegraphics[width=0.95\textwidth,angle=270]{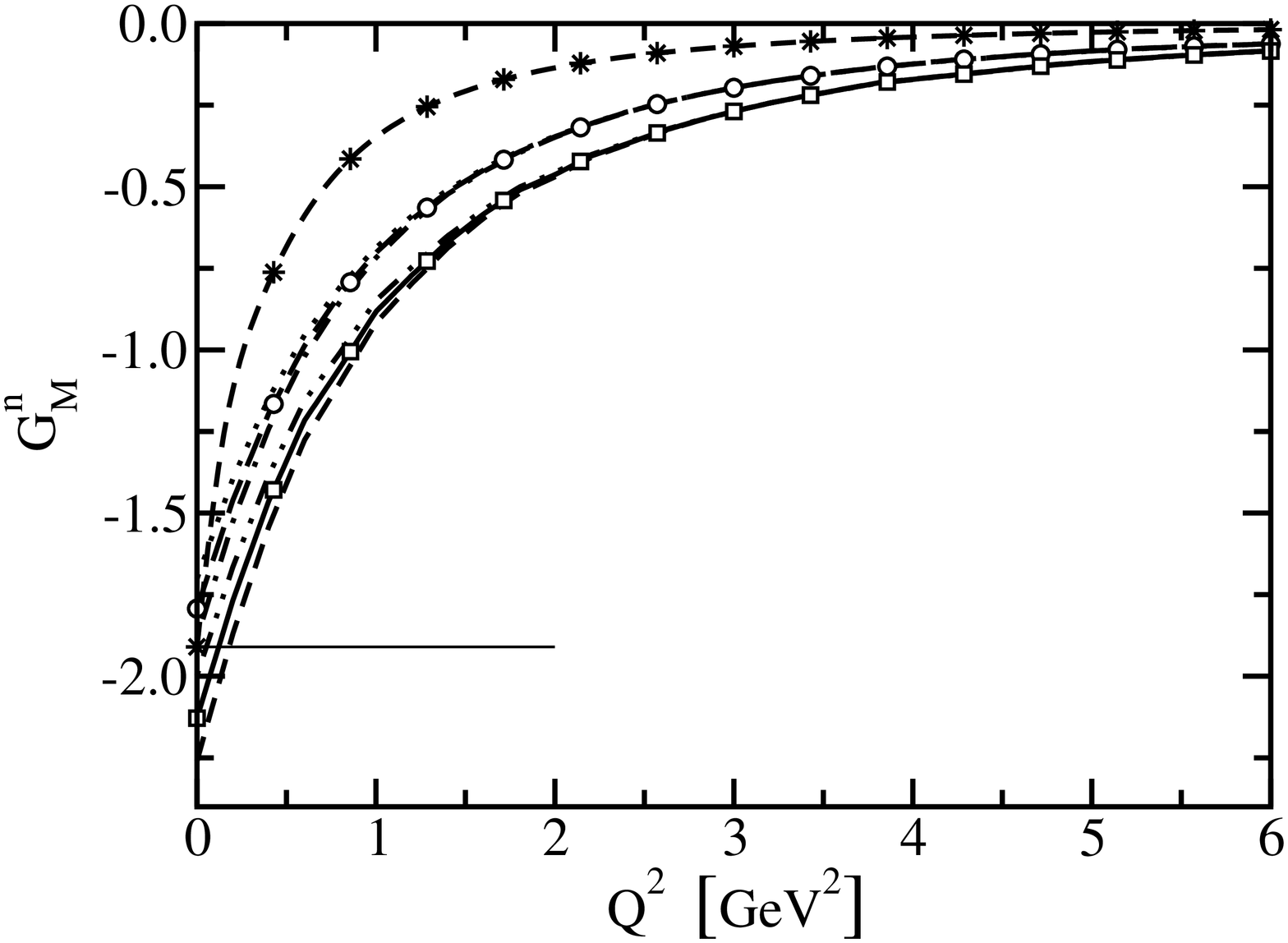}%
\hspace*{1em}}
\end{minipage}
\caption{\label{plot1}
Response of nucleon form factors to variations in the magnetic moment of the axial-vector diquark: $\mu_{1^+}=1,2,3$; with $\chi_{1^+}=1$, $\kappa_{\cal T}=2$.  
%\textit{Left Column} -- $G_E^p(Q^2)$, upper entry; $G_E^n(Q^2)$, lower entry.  \textit{Right Column} -- $G_M^p(Q^2)$, upper entry; $G_M^n(Q^2)$, lower entry; and in both panels the horizontal line marks the experimental value of the magnetic moment.  
The legend in the top-left panel applies to all; the dashed-line marked by ``$\ast$'' is a fit to experimental data \protect\cite{Walcher03} and the dashed-line marked by ``$+$'' in the lower-left panel is the fit to $G_E^n(Q^2)$ of Ref.\,\cite{galster}; and the horizontal lines in the right panels mark the experimental value of the nucleon's magnetic moment.}
\end{figure}

\begin{figure}[t]
\begin{minipage}{0.45\textwidth}
\centerline{\hspace*{2.5em}%
\includegraphics[width=0.95\textwidth,angle=270]{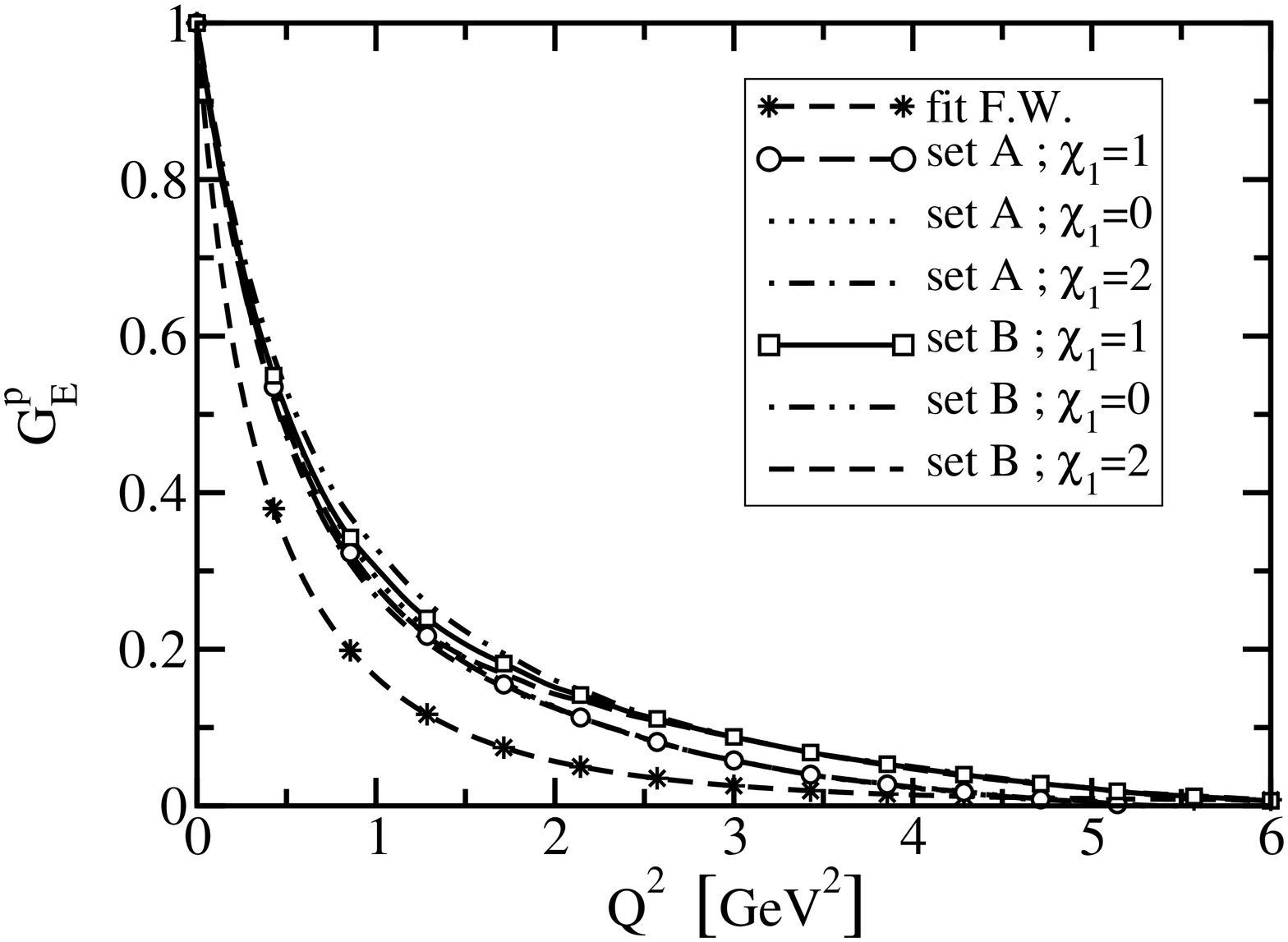}}
\end{minipage}
\hfill
\begin{minipage}{0.45\textwidth}
\centerline{\includegraphics[width=0.95\textwidth,angle=270]{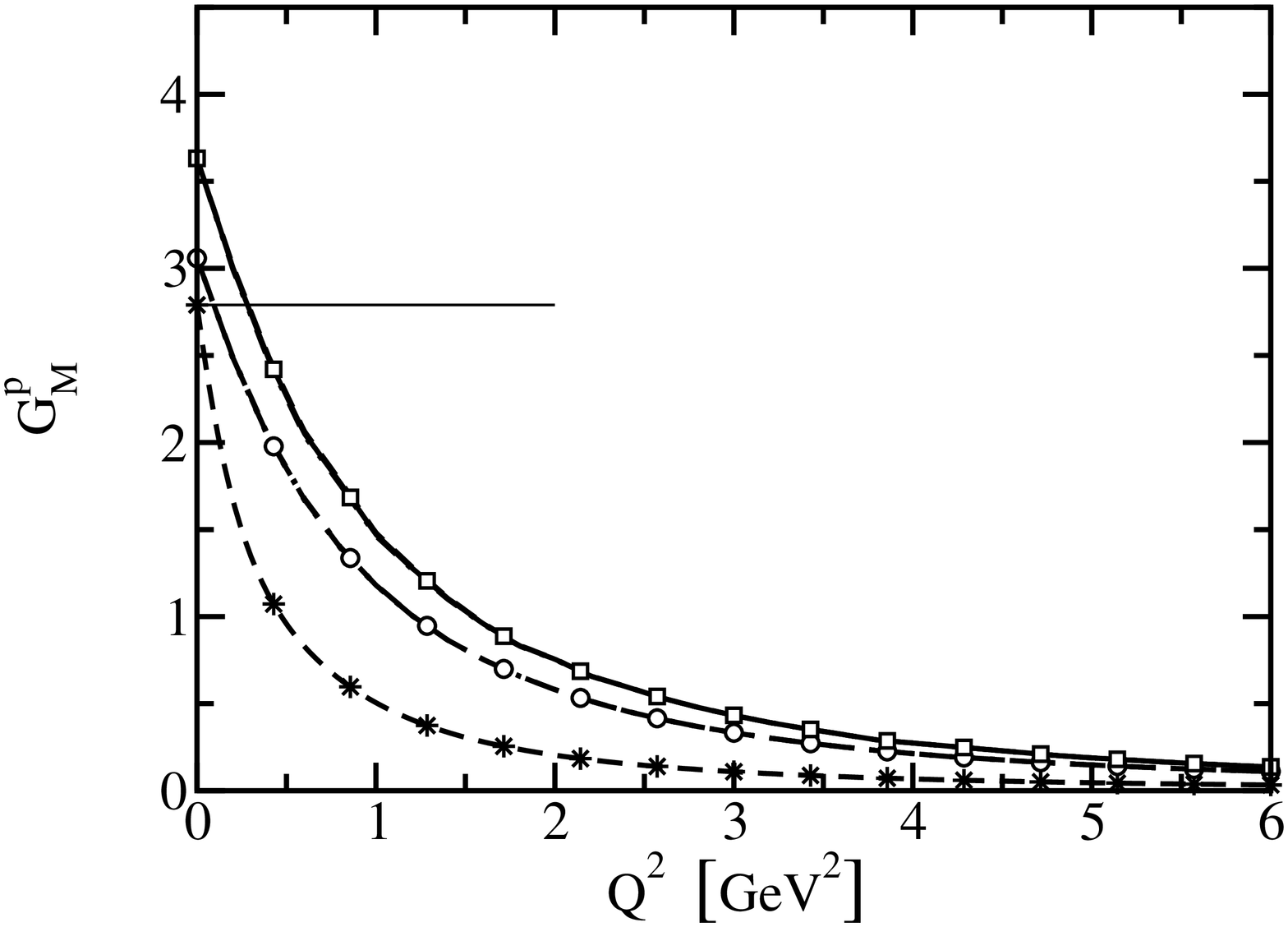}%
\hspace{1em}}
\end{minipage}
\\
\begin{minipage}{0.45\textwidth}
\centerline{\hspace*{2.5em}%
\includegraphics[width=0.95\textwidth,angle=270]{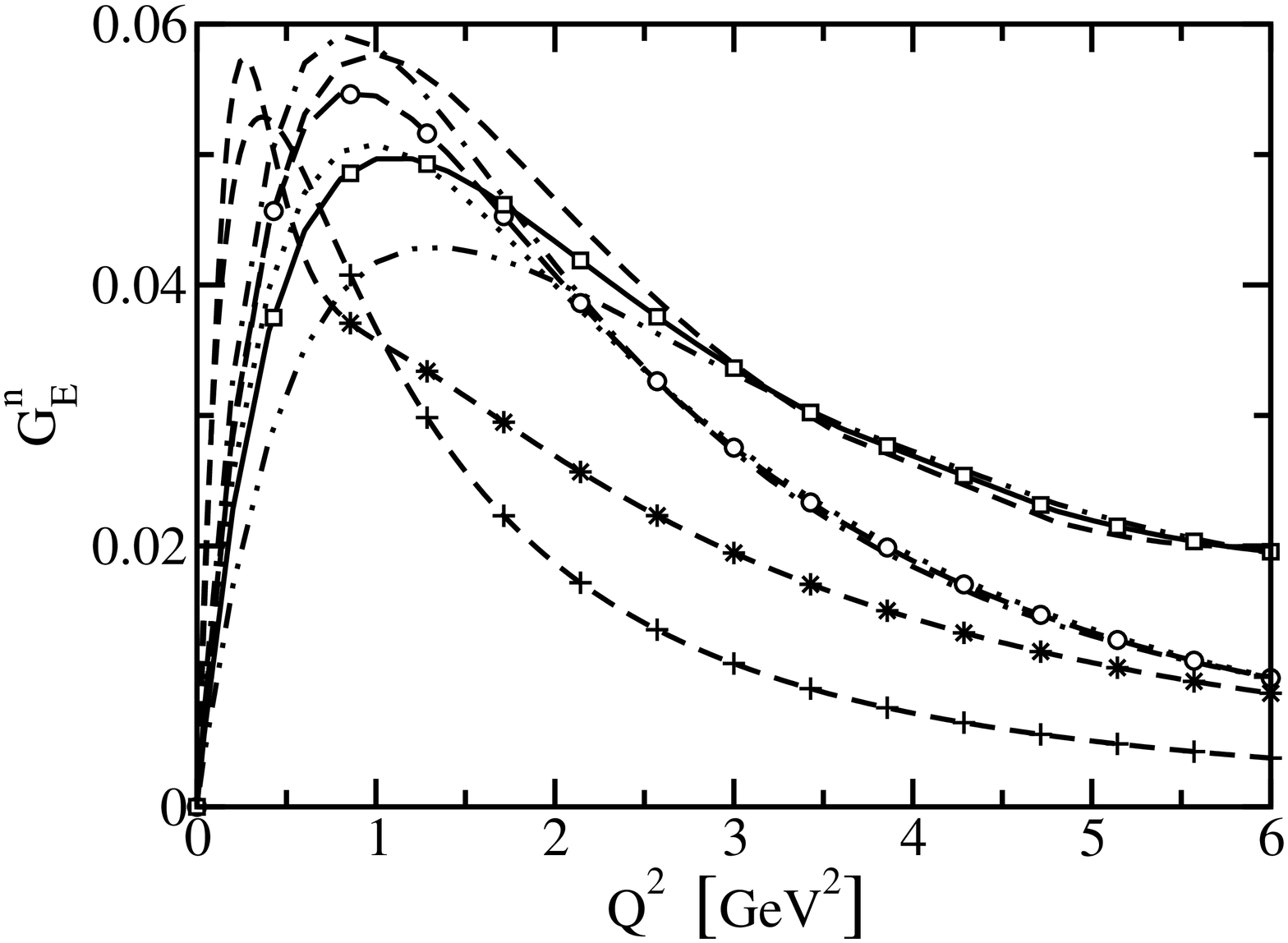}}
\end{minipage}
\hfill
\begin{minipage}{0.45\textwidth}
\centerline{\includegraphics[width=0.95\textwidth,angle=270]{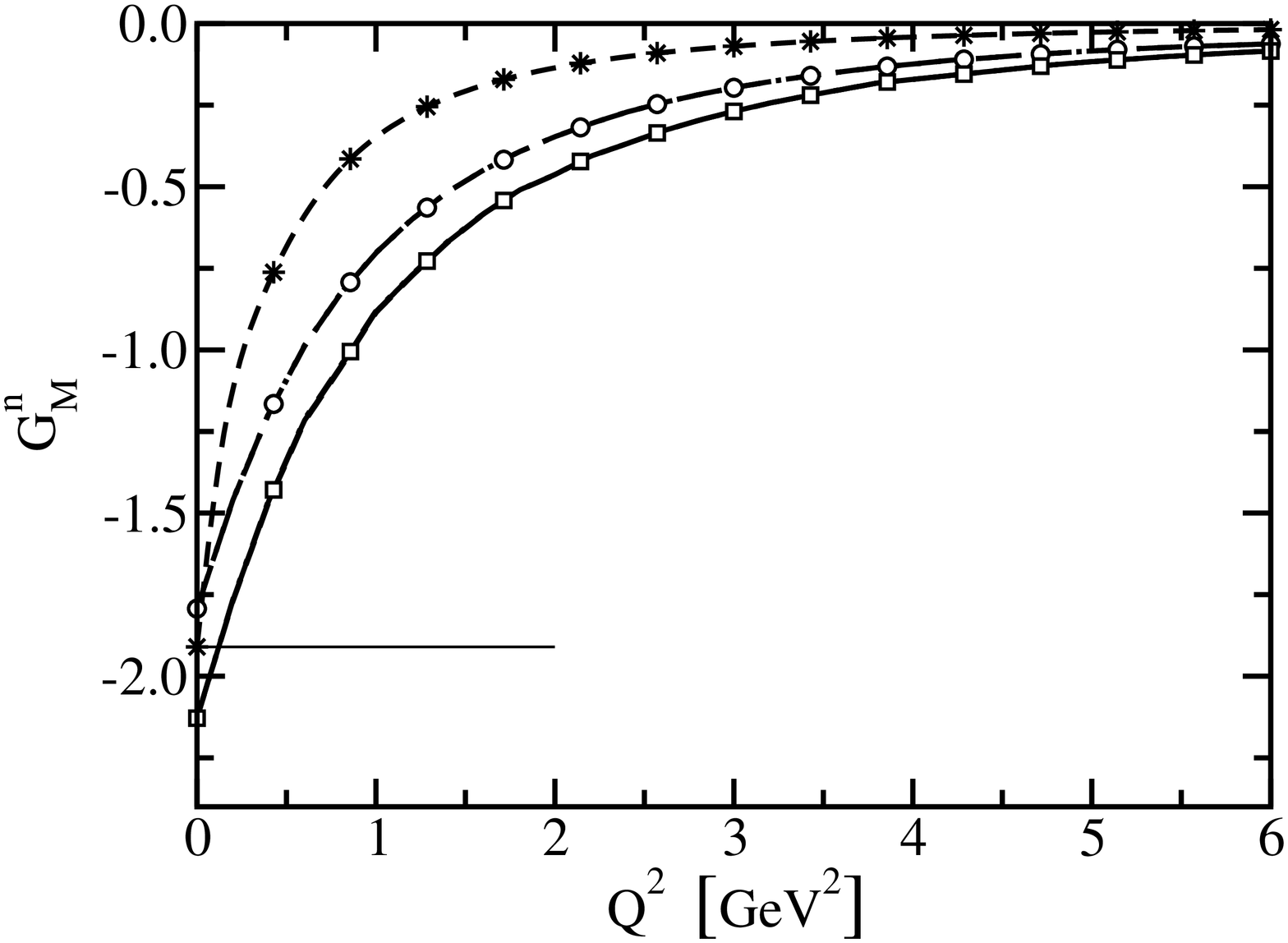}%
\hspace*{1em}}
\end{minipage}
\caption{\label{plot2}
Response of nucleon form factors to variations in the electric quadrupole moment of the axial-vector diquark: $\chi_{1^+}=0,1,2$; with $\mu_{1^+}=2$, $\kappa_{\cal T}=2$.  The other features are described in the caption of Fig.\,\protect\ref{plot1}.
%horizontal lines in the right panels mark the experimental value of the nucleon's magnetic moment.  In all panels, the dashed curve is a fit to experimental data \protect\cite{Walcher03}, while in the lower left panel the dash-dot curve is the fit of Ref.\,\cite{galster}.
%\textit{Left Column} -- $G_E^p(Q^2)$, upper entry; $G_E^n(Q^2)$, lower entry.  \textit{Right Column} -- $G_M^p(Q^2)$, upper entry; $G_M^n(Q^2)$, lower entry; and in both panels the horizontal line marks the experimental value of the magnetic moment.  In all panels, the dashed curve is a fit to experimental data \protect\cite{Walcher03}, while in the lower left panel the dash-dot curve is the fit of Ref.\,\cite{galster}.
}
\end{figure}

\begin{figure}[t]
\begin{minipage}{0.45\textwidth}
\centerline{\hspace*{2.5em}%
\includegraphics[width=0.95\textwidth,angle=270]{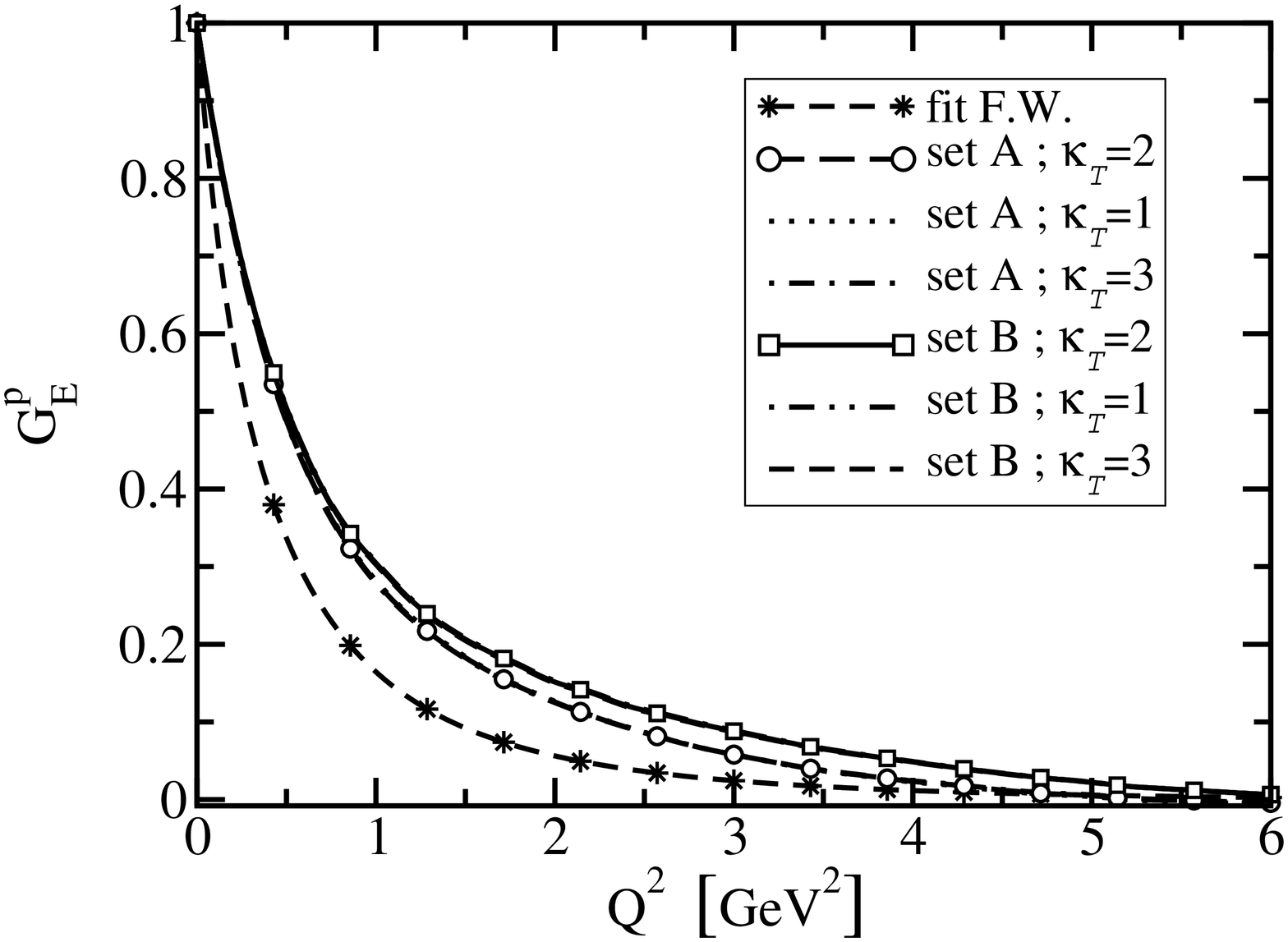}}
\end{minipage}
\hfill
\begin{minipage}{0.45\textwidth}
\centerline{\includegraphics[width=0.95\textwidth,angle=270]{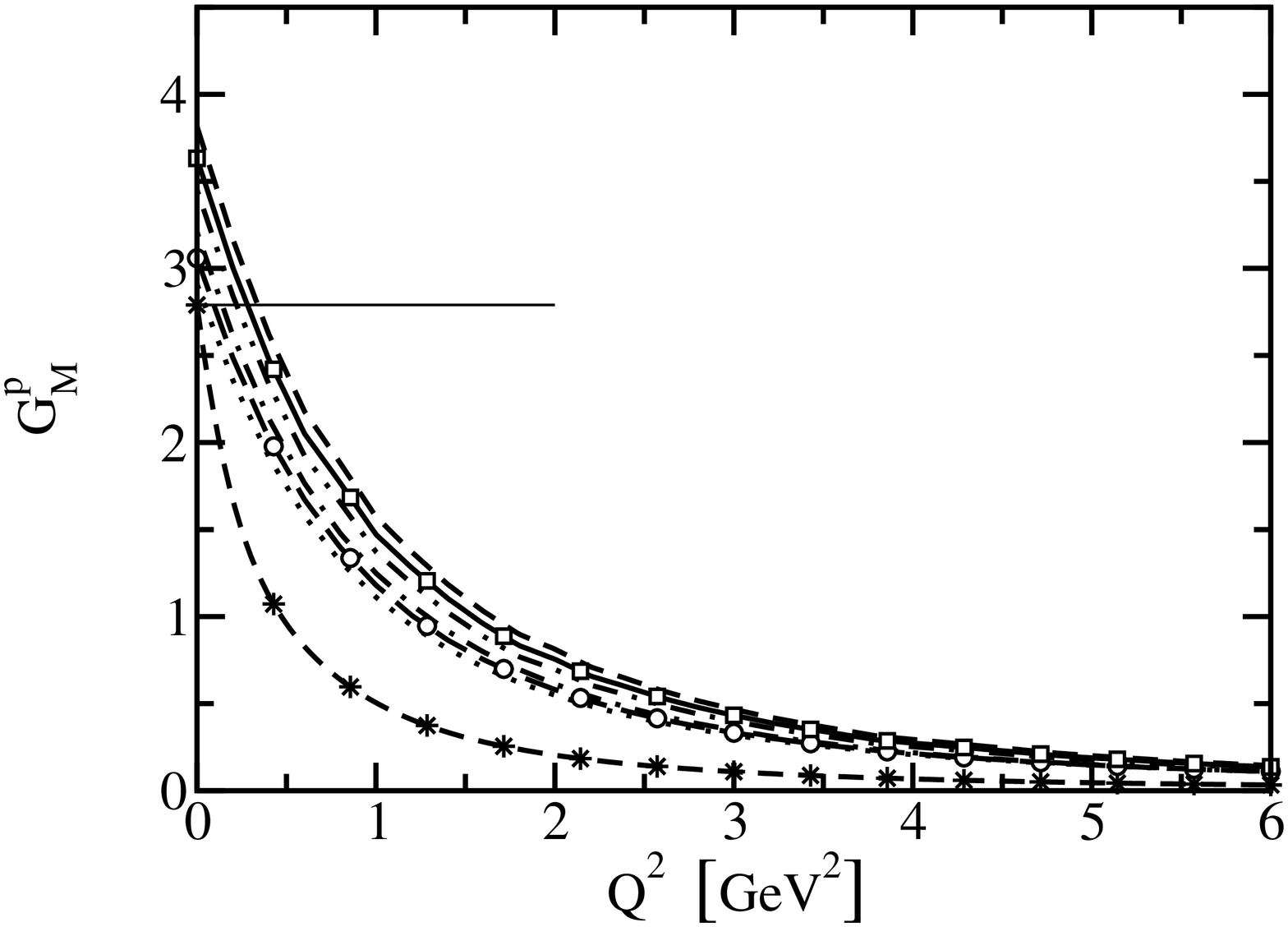}%
\hspace{1em}}
\end{minipage}
\\
\begin{minipage}{0.45\textwidth}
\centerline{\hspace*{2.5em}%
\includegraphics[width=0.95\textwidth,angle=270]{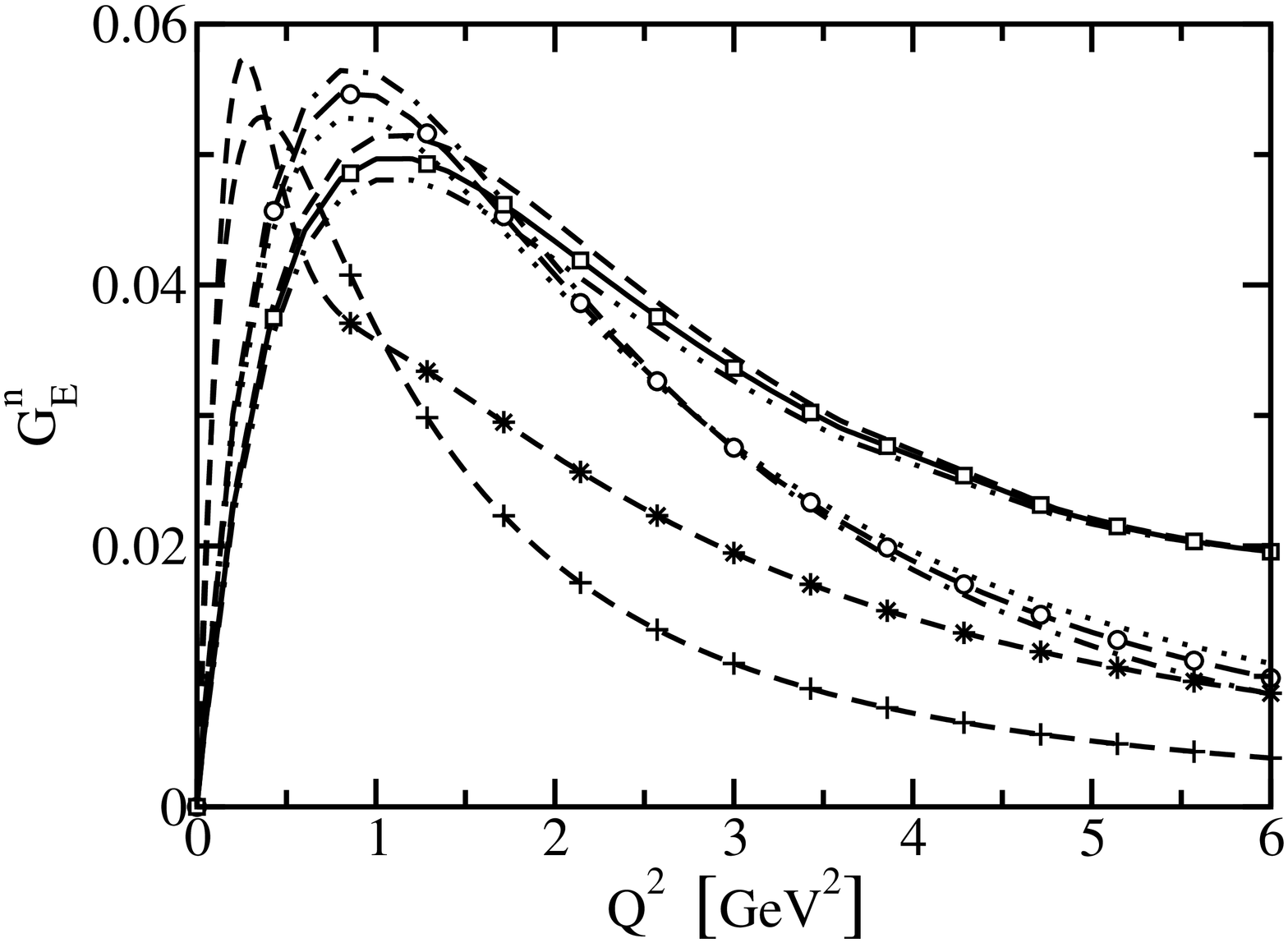}}
\end{minipage}
\hfill
\begin{minipage}{0.45\textwidth}
\centerline{\includegraphics[width=0.95\textwidth,angle=270]{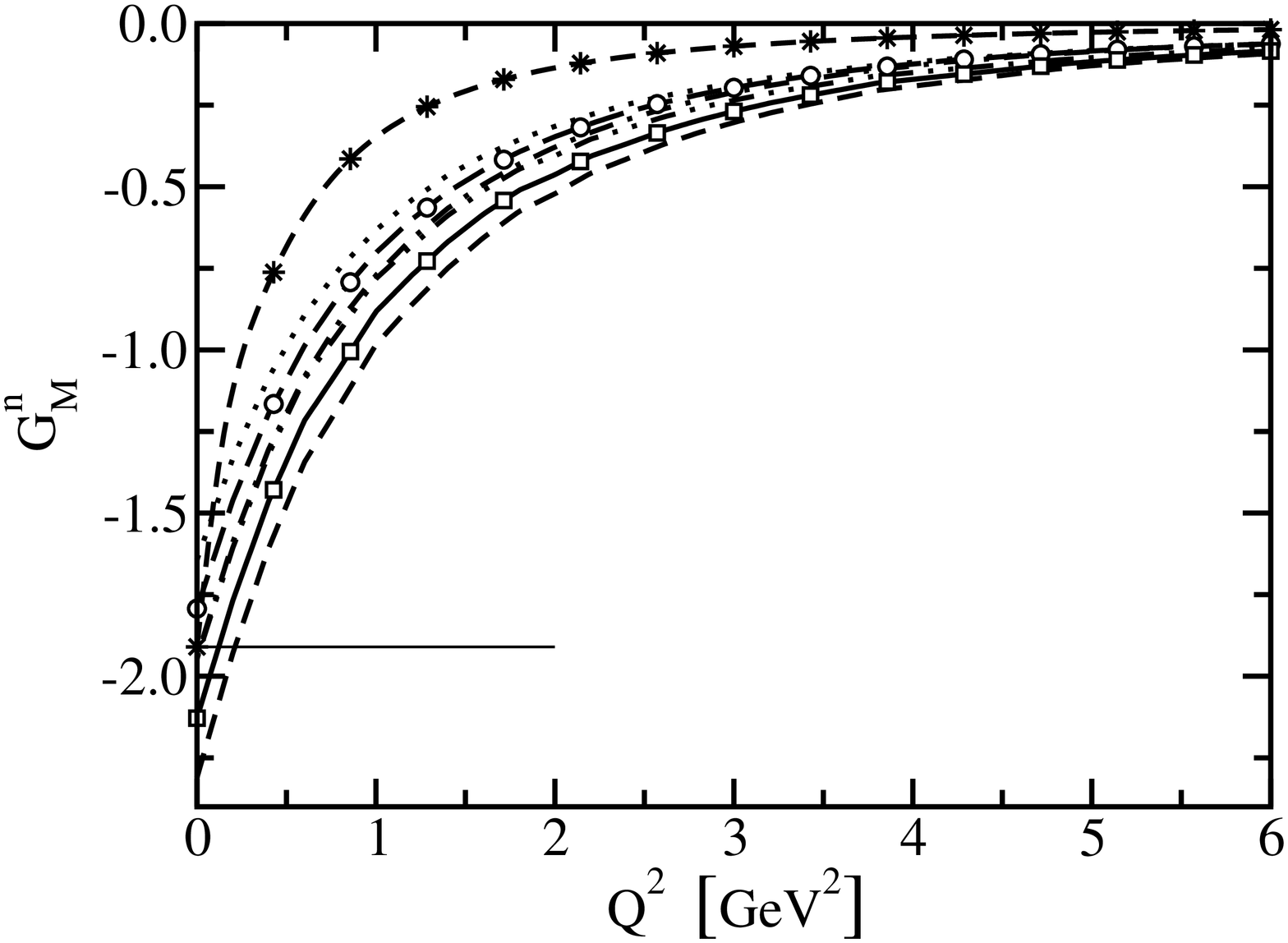}%
\hspace*{1em}}
\end{minipage}
\caption{\label{plot3}
Response of nucleon form factors to variations in the strength of the electromagnetic axial-vector $\leftrightarrow$ scalar diquark transition: $\kappa_{\cal T}=1,2,3$; with $\mu_{1^+}=2$, $\chi_{1^+}=1$.  The other features are described in the caption of Fig.\,\protect\ref{plot1}.
%
%The horizontal lines in the right panels mark the experimental value of the nucleon magnetic's moment.  In all panels, the dashed curve is a fit to experimental data \protect\cite{Walcher03}, while in the lower left panel the dash-dot curve is the fit of Ref.\,\cite{galster}.
%\textit{Left Column} -- $G_E^p(Q^2)$, upper entry; $G_E^n(Q^2)$, lower entry.  \textit{Right Column} -- $G_M^p(Q^2)$, upper entry; $G_M^n(Q^2)$, lower entry; and in both panels the horizontal line marks the experimental value of the magnetic moment.  In all panels, the dashed curve is a fit to experimental data \protect\cite{Walcher03}, while in the lower left panel the dash-dot curve is the fit of Ref.\,\cite{galster}.
}
\end{figure}

Nucleon electromagnetic form factors associated with the tabulated values of static properties are presented in Figs.\,\ref{plot1}--\ref{plot3}.  These figures confirm and augment the information in Tables~\ref{radii}--\ref{moments}.  Consider, e.g., the electric form factors.  One observes that the differences between results obtained with Set~A and Set~B generally outweigh those delivered by variations in the parameters characterising the axial-vector diquark's electromagnetic properties.  The proton's electric form factor, in particular, is largely insensitive to these parameters, and it is apparent that the nucleon's electric form factor only responds notably to variations in $\chi_{1^+}$, Fig.\,\ref{plot2}.  The nucleons' magnetic form factors exhibit the greatest sensitivity to the axial-vector diquarks's electromagnetic properties but in this case, too, the differences between Set~A and Set~B are more significant.  For $Q^2 \gtrsim 4\,$GeV$^2$ there is no sensitivity to the diquarks' electromagnetic parameters in any curve.  This is naturally because our parametrisation expresses the perturbative limit, Eq.\,(\ref{pQCDavdq}).  It is thus apparent from these figures that the behaviour of the nucleons' form factors is primarily determined by the information encoded in the Faddeev amplitudes.

Our results show that the nucleons' electromagnetic properties are sensitive to the strength of axial-vector diquark correlations in the bound state and react to the electromagnetic properties of these correlations.  In all cases the dependence is readily understood intuitively.  However, taken together our results indicate that one cannot readily tune the model's parameters to provide a uniformly good account of nucleon properties: something more than dressed-quark and -diquark degrees of freedom is required.

\subsubsection{Chiral corrections}
It is appropriate now to examine effects that arise through coupling to pseudoscalar mesons.  As with baryon masses, there are two types of contributions from meson loops to electromagnetic form factors: regularisation-scheme-dependent terms, which are analytic in the neighbourhood of $\hat m = 0$; and nonanalytic scheme-independent terms.  For the static properties presented in Tables~\ref{radii}--\ref{moments}, the leading-order scheme-independent contributions are \cite{kubis}
\begin{eqnarray}
\label{rpnpion}
\langle r_{p\atop n}^2\rangle^{1-loop}_{NA} &=& \mp\,\frac{1+5 g_A^2}{32 \pi^2 f_\pi^2} \,\ln (\frac{m_\pi^2}{M_N^2}) \,, \\
\label{rpnmpion}
\langle (r_{N}^\mu)^2\rangle^{1-loop}_{NA} &=& -\,\frac{1+5 g_A^2}{32 \pi^2 f_\pi^2} \,\ln (\frac{m_\pi^2}{M_N^2})+ \frac{g_A^2\, M_N}{16 \pi f_\pi^2 \mu_v} \frac{1}{m_\pi} \,,\\
\label{mpnpion}
(\mu_{p\atop n})_{NA}^{1-loop} & = & \mp \, \frac{g_A^2\, M_N}{4\pi^2 f_\pi^2}\, m_\pi\,,
\end{eqnarray}
where $g_A=1.26$, $f_\pi=0.0924$\,GeV\,$=1/(2.13 \,{\rm fm})$, $\mu_v=\mu_p-\mu_n$.  Clearly, the radii diverge in the chiral limit, a much touted aspect of chiral corrections.  

While these scheme-independent terms are immutable, at physical values of the pseudoscalar meson masses they do not usually provide the dominant contribution to observables: that is provided by the regularisation-parameter-dependent terms.  This is apparent for baryon masses in Ref.\,\cite{hechtfe} and for the pion charge radius in Ref.\,\cite{bender}.  It is particularly important here, as is made plain by a consideration of the neutron charge radius.  From Eq.\,(\ref{rpnpion}), one obtains
\begin{equation}
\langle r_{n}^2\rangle^{1-loop}_{NA} = - \,(0.48\,{\rm fm})^2\,,
\end{equation}
which is more than twice the experimental value.  On the other hand, the contribution from the low energy constants is \cite{kubis}
\begin{equation}
\langle r_{n}^2\rangle^{1-loop}_{lec} = +\, (0.69\,{\rm fm})^2\,,
\end{equation}
which is four-times larger in magnitude than the experimental value and has the opposite sign.  This emphasises the delicate cancellation that is arranged in chiral perturbation theory to fit the neutron's charge radius.  In this instance the remaining important piece is the neutron's Foldy term: $3 \mu_n/(2 M_n^2)=-(0.35\,{\rm fm})^2$, which is also a fitted quantity.  Moreover, for the magnetic radii it is established that, at the physical pion mass, the leading chiral limit behaviour is not a good approximation \cite{kubis}.  Additional discussion of issues that arise in formulating a chiral expansion in the baryon sector and its convergence may be found in Ref.\,\cite{fuchs}. 

Since regularisation-parameter-dependent parts of the chiral loops are important we follow Ref.\,\cite{ashley} and estimate the corrections using modified formulae that incorporate a single parameter which  mimics the effect of regularising the integrals.  Thus Eqs.\,(\ref{rpnpion}) \& (\ref{rpnmpion}) are rewritten
\begin{eqnarray}
\label{rpnpionR}
\langle r_{p\atop n}^2\rangle^{1-loop^R}_{NA} &=& \mp\,\frac{1+5 g_A^2}{32 \pi^2 f_\pi^2} \,\ln (\frac{m_\pi^2}{m_\pi^2+\lambda^2}) \,, \\
\nonumber \langle (r_{N}^\mu)^2\rangle^{1-loop^R}_{NA} &=& -\,\frac{1+5 g_A^2}{32 \pi^2 f_\pi^2} \,\ln (\frac{m_\pi^2}{m_\pi^2+\lambda^2}) + \frac{g_A^2\, M_N}{16 \pi f_\pi^2 \mu_v} \frac{1}{m_\pi} \,
\frac{2}{\pi}\arctan(\frac{\lambda}{m_\pi})\,,\\
\label{rpnmpionR}\\
\label{mpnpionR}
(\mu_{p\atop n})_{NA}^{1-loop^R}& = & \mp \, \frac{g_A^2\, M_N}{4\pi^2 f_\pi^2}\, m_\pi\, \frac{2}{\pi}\arctan(\frac{\lambda^3}{m_\pi^3})\,,
\end{eqnarray}
wherein $\lambda$ is a regularisation mass-scale, for which a typical value is $\sim 0.4\,$GeV \cite{ashley}.  NB.\ The loop contributions vanish when the pion mass is much larger than the regularisation scale, as required: very massive states must decouple from low-energy phenomena.  

\begin{table}[bt]
\begin{center}
\caption{\label{picorrected} Row~1 -- static properties calculated with Set~B diquark masses, Table~\protect\ref{ParaFix}, and $\mu_{1^+}=2$, $\chi_{1^+}=1$, $\kappa_{\cal T}=2$: charge radii in fm, with $r_n$ defined in Table~\ref{radii}; and magnetic moments in nuclear magnetons.  Row~2 adds the corrections of Eqs.\,(\protect\ref{rpnpionR})--(\protect\ref{mpnpionR}) with $\lambda=0.3\,$GeV.  $\varsigma$ in row~$n$, is the rms relative-difference between the entries in row $n$ and 3.}
\begin{tabular*}{1.0\textwidth}{c@{\extracolsep{0ptplus1fil}}
|c@{\extracolsep{0ptplus1fil}}
c@{\extracolsep{0ptplus1fil}}c@{\extracolsep{0ptplus1fil}} 
c@{\extracolsep{0ptplus1fil}}c@{\extracolsep{0ptplus1fil}}
c@{\extracolsep{0ptplus1fil}}c@{\extracolsep{0ptplus1fil}}
|c@{\extracolsep{0ptplus1fil}}}
  & $r_p$ & $r_n$ & $r_p^\mu$ & $r_n^\mu$ & $\mu_p$ & $-\mu_n$ && $\varsigma$\\\hline
$q$-$(qq)$ core & 0.595 & 0.169 & 0.449 & 0.449 & 3.63 & 2.13 && 0.39\\
$+\pi$-loop correction & 0.762 & 0.506 & 0.761 & 0.761 & 3.05 & 1.55 && 0.23 \\\hline
experiment & 0.847 & 0.336 & 0.836 & 0.889 & 2.79& 1.91 && \\\hline
\end{tabular*}
\end{center}
\end{table}

We return now to the calculated values of the nucleons' static properties,  Tables~\ref{radii}--\ref{moments}, and focus on the Set~B results obtained with $\mu_{1^+}=2$, $\chi_{1^+}=1$, $\kappa_{\cal T}=2$.  Recall that Set~B was chosen to give inflated values of the nucleon and $\Delta$ masses in order to make room for chiral corrections, and therefore one may consistently apply the corrections in Eqs.\,(\ref{mpnpion}), (\ref{rpnpionR}) \& (\ref{rpnmpionR}) to the static properties.  With $\lambda=0.3\,$GeV this yields the second row in Table~\ref{picorrected}: the regularised chiral corrections reduce the rms relative-difference signficantly.  This crude analysis, complementing Ref.\,\cite{hechtfe}, suggests that a veracious description of baryons can be obtained using dressed-quark and -diquark degrees of freedom augmented by a sensibly regulated pseudoscalar meson cloud.

\begin{figure}[t]
\centerline{%\hspace*{2em}%
\includegraphics[width=0.75\textwidth,angle=270]{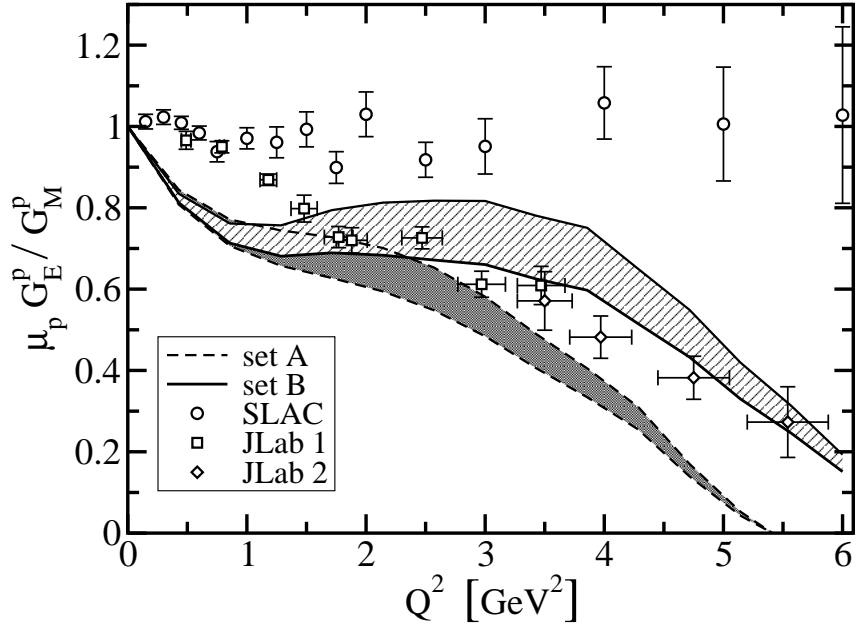}}
\caption{\label{plotGEpGMp} Proton form factor ratio: $\mu_p\, G_E^p(Q^2)/G_M^p(Q^2)$.  Calculated results: \textit{lower band} - Set~A in Table~\ref{ParaFix}; and \textit{upper band} - Set~B.  For both bands, $G_E^p(Q^2)$ was calculated using the point-particle values: $\mu_{1^+}=2$ \& $\chi_{1^+}=1$, Eq.\,(\protect\ref{pointp}), and $\kappa_{\cal T} = 2$, Eq.\,(\ref{kTbest}); i.e., the reference values in Tables~\protect\ref{radii}--\protect\ref{moments}.  Variations in the axial-vector diquark parameters used to evaluate $G_E^p(Q^2)$ have little effect on the plotted results.  The width of the bands reflects the variation in $G_M^p(Q^2)$ with axial-vector diquark parameters and, in both cases, the upper border is obtained with $\mu_{1^+}=3$, $\chi_{1^+}=1$ and $\kappa_{\cal T}= 2$, while the lower has $\mu_{1^+}= 1$.  The data are: \textit{squares} - Ref.\,\cite{jones}; \textit{diamonds} - Ref.\,\cite{gayou}; and \textit{circles} - Ref.\,\cite{walker}.}
\end{figure}

\subsubsection{Form factor ratios}
In Fig.\,\ref{plotGEpGMp} we plot the ratio $\mu_p\, G_E^p(Q^2)/G_M^p(Q^2)$.  The behaviour of the experimental data at small $Q^2$ is readily understood.  In the neighbourhood of $Q^2=0$, 
\begin{equation}
\mu_p\,\frac{ G_E^p(Q^2)}{G_M^p(Q^2)} = 1 - \frac{Q^2}{6} \,\left[ (r_p)^2 - (r_p^\mu)^2 \right]\,,
\end{equation}
and because $r_p\approx r_p^\mu$ the ratio varies by less than 10\% on $0<Q^2< 0.6\,$GeV$^2$, if the form factors are approximately dipole.  This is evidently true of the experimental data.  In our calculation, without chiral corrections, Table~\ref{radii} \& \ref{magradii}, $r_p> r_p^\mu$.  Hence the ratio must fall immediately with increasing $Q^2$.  Incorporating pion loops, we obtain the results in Row~2 of Table\,\ref{picorrected}, which have $r_p\approx r_p^\mu$.  The small $Q^2$ behaviour of this ratio is thus materially affected by the proton's pion cloud.

True pseudoscalar mesons are not pointlike and therefore pion cloud contributions to form factors diminish in magnitude with increasing $Q^2$.  For example, in a study of the $\gamma N \to \Delta$ transition \cite{sato}, pion cloud contributions to the M1 form factor fall from 50\% of the total at $Q^2=0$ to $\lsim 10$\% for $Q^2\gtrsim 2\,$GeV$^2$.  Hence, the evolution of $\mu_p\, G_E^p(Q^2)/G_M^p(Q^2)$ on $Q^2\gtrsim 2\,$GeV$^2$ is primarily determined by the quark core of the proton.  This is evident in Fig.\,\ref{plotGEpGMp}, which illustrates that, on $Q^2\in (1,5)\,$GeV$^2$, $\mu_p\, G_E^p(Q^2)/G_M^p(Q^2)$ is sensitive to the parameters defining the axial-vector-diquark--photon vertex.  The response diminishes with increasing $Q^2$ because our parametrisation expresses the perturbative limit, Eq.\,(\ref{pQCDavdq}).  

The behaviour of $\mu_p\, G_E^p(Q^2)/G_M^p(Q^2)$ on $Q^2\gtrsim 2\,$GeV$^2$ is determined either by correlations expressed in the Faddeev amplitude, the electromagnetic properties of the constituent degrees of freedom, or both.  The issue is decided by the fact that the magnitude and trend of the results are not materially affected by the axial-vector-diquarks' electromagnetic parameters.  This observation suggests strongly that the ratio's evolution is due primarily to spin-isospin correlations in the nucleon's Faddeev amplitude.  One might question this conclusion, and argue instead that the difference between the results depicted for Set~A cf.\ Set~B originates in the larger quark-core nucleon mass obtained with Set~B (Table~\ref{ParaFix}) which affects $G_E^p(Q^2)$ through Eq.\,(\ref{GEpeq}).  We checked and that is not the case.  Beginning with the Set~B results for $F_{1,2}$ we calculated the electric form factor using the Set~A nucleon mass and then formed the ratio.  The result is very different from the internally consistent Set~A band in Fig.\,\ref{plotGEpGMp}; e.g., it drops more steeply and lies uniformly below, and crosses zero for $Q^2\approx 3.7\,$GeV$^2$.  

It is noteworthy that Set~B, which anticipates pion cloud effects, is in reasonable agreement with both the trend and magnitude of the polarisation transfer data \cite{jones,roygayou,gayou}.  NB.\ Neither this nor the Rosenbluth \cite{walker} data played any role in the preceding analysis or discussion.  

\begin{figure}[t]
\centerline{\hspace*{2em}%
\includegraphics[width=0.75\textwidth,angle=270]{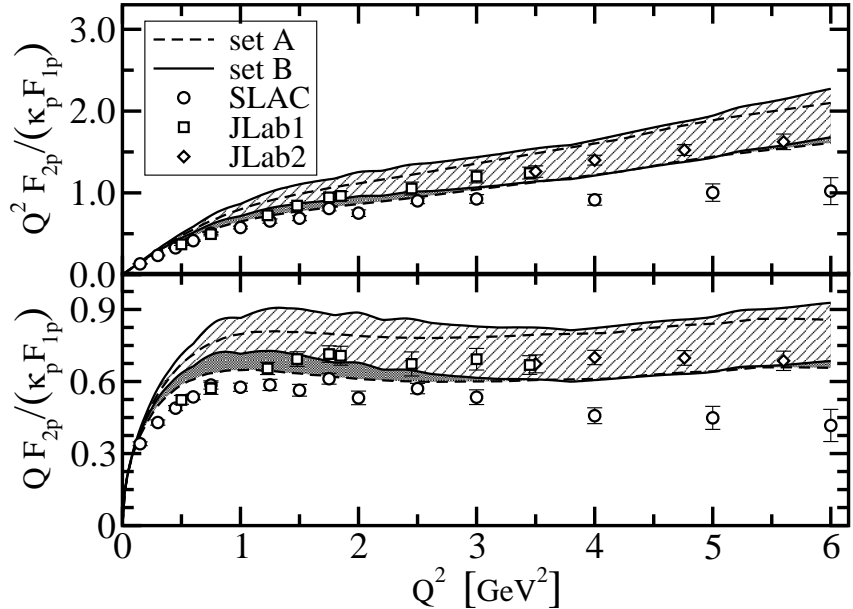}}
\caption{\label{plotF2F1} Proton Pauli$/$Dirac form factor ratios.  The data are as described in Fig.\,\protect\ref{plotGEpGMp}, as are the bands except that here the upper border is obtained with $\mu_{1^+}=1$, $\chi_{1^+}=1$ and $\kappa_{\cal T}= 2$, and the lower with $\mu_{1^+}=3$.}
\end{figure}

We have also examined the proton's Dirac and Pauli form factors in isolation.  On the domain covered, neither $F_1(Q^2)$ nor $F_2(Q^2)$ show any sign they have achieved the asymptotic behaviour anticipated from perturbative QCD.  

In Fig.\,\ref{plotF2F1} we depict weighted ratios of these form factors.  
%The similarity between $F_{1,2}$ calculated with the parameters of Set~A and Set~B is evident.  
Our numerical results are consistent with 
\begin{equation}
 \sqrt{Q^2}\,\frac{ F_2(Q^2)}{F_1(Q^2)}\; 
 %\stackrel{Q^2\gtrsim 2\,{\rm GeV}^2}{\approx}\; 
 \approx {\rm constant},\; 2 \lesssim  Q^2 ({\rm GeV}^2) \lesssim 6\,,
\end{equation}
as are the polarisation transfer data.  Such behaviour has been argued to indicate the presence of substantial orbital angular momentum in the proton \cite{millerfrank,ralstonjain}.  Orbital angular momentum is not a Poincar\'e invariant.  However, if absent in a particular frame, it will almost inevitably appear in another frame related via a Poincar\'e transformation.  Nonzero quark orbital angular momentum is the necessary outcome of a Poincar\'e covariant description.  This is why the covariant Faddeev amplitude describing a nucleon is a matrix-valued function with a rich structure that, in the nucleons' rest frame, corresponds to a relativistic wave function with $s$-wave, $p$-wave and even $d$-wave components. (Details can be found in Ref.\,\cite{oettelthesis}, Sec.~2.4.)  The result in Fig.\,\ref{plotF2F1} is not significantly influenced by details of the diquarks' electromagnetic properties.  Instead, the behaviour is primarily governed by correlations expressed in the proton's Faddeev amplitude and, in particular, by the amount of intrinsic quark orbital angular momentum \cite{blochff}.  NB.\ 
%is not new nor surprising.  It 
This phenomenon is analogous to that observed in connection with the pion's electromagnetic form factor.  In that instance axial-vector components of the pion's Bethe-Salpeter amplitude are responsible for the large $Q^2$ behaviour of the form factor: they alone ensure $Q^2 F_\pi(Q^2) \approx \,$constant for truly ultraviolet momenta \cite{mrpion}.  These components are required by covariance \cite{mrt98} and signal the presence of quark orbital angular momentum in the pseudoscalar pion.

\begin{figure}[t]
\centerline{\hspace*{2em}%
\includegraphics[width=0.75\textwidth,angle=270]{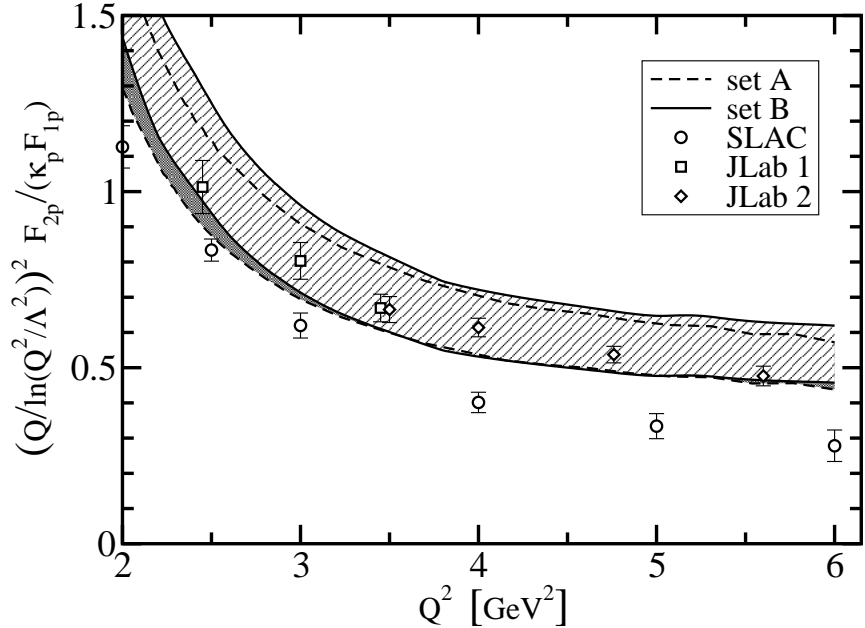}}
\caption{\label{plotF2F1log}  Weighted proton Pauli$/$Dirac form factor ratio, calculated with $\Lambda= M_N=0.94\,$GeV.  The bands are as described in Fig.\,\protect\ref{plotF2F1}, as are the data.}
\end{figure}

In Fig.\,\ref{plotF2F1log} we plot another weighted ratio of Pauli and Dirac form factors.  A perturbative QCD analysis \cite{belitsky} that considers effects arising from both the proton's leading- and subleading-twist light-cone wave functions, the latter of which represents quarks with one unit of orbital angular momentum, suggests
\begin{equation}
\label{scaling}
\frac{Q^2}{[\ln Q^2/\Lambda^2]^2} \, \frac{F_2(Q^2)}{F_1(Q^2)} =\,{\rm constant,}\;\; Q^2\gg \Lambda^2\,,
\end{equation}  
where $\Lambda$ is a mass-scale that corresponds to an upper-bound on the domain of nonperturbative (soft) momenta.  This scaling hypothesis is not predictive unless the value of $\Lambda$ is known \textit{a priori}.  However, $\Lambda$ cannot be computed in perturbation theory.  

A scale of this type is not an elemental input to our calculation.  It is instead a derivative quantity that expresses the net integrated effect of many basic features, among which are the mass-scale characterising quark-dressing and that implicit in the support of the Faddeev amplitude.  Extending our calculation to larger $Q^2$ is not a problem in principle, and the numerical challenge may readily be met by choosing to use something more powerful than a single desktop computer.  With such information we would be confident of credibly exploring the validity of Eq.\,(\ref{scaling}); lacking it, we can only provide a plausible argument.  

One needs an estimate of a reasonable value for $\Lambda$.  The nucleon's mass, $M_N$, is one natural mass-scale in our calculation.  Other relevant mass-scales are those which characterise the electromagnetic size of the nucleon and its constituents.  A dipole mass-scale for the proton is approximately $0.85\,$GeV; the dressed-quark-photon-vertex is characterised by a monopole mass-scale of $0.8\,$GeV \cite{blochff}; and the diquark-photon vertices, by monopole mass-scales $\sqrt{3} \, m_{J^P} \approx 1.0\,$-$\,1.5\,$GeV, Eqs.\,(\ref{Gamma0plus}) \& (\ref{AnsatzF1}).  

As an adjunct one can consider the dressed-quark mass function, defined in Eq.\,(\ref{ZMdef}) and discussed thereabout.  A nonzero mass function in the chiral limit is an essentially nonperturbative phenomenon.  Hence the ratio 
\begin{equation}
R^0_u(Q^2):= \frac{M_{\hat m=0}(Q^2)}{M_{\hat m_u}(Q^2)}
\end{equation}
vanishes on the perturbative-$Q^2$ domain.  For $Q^2=0$, on the other hand, calculations typically yield \cite{wrightgap} $R^0_{u}(0)= 0.96$; i.e., the mass function's behaviour is almost completely nonperturbative.  The $Q^2$-evolution of $R^0_{u}(Q^2)$ can therefore guide in demarcating the nonperturbative domain.  Reference~\cite{bhagwat} provides a mass-function that agrees pointwise with quenched-QCD lattice data and gives a unique chiral-limit mass function.  From these results one finds that $R^0_{u}(Q^2> 4 M_N^2) <0.5\,R^0_{u}(0)$; viz., perturbative effects are dominant in the $u$-quark mass function on $Q^2\in (4 M_N^2,\infty)$.  On the other hand, $0.8 < R^0_{u}(Q^2<M_N^2)$ and hence the $u$-quark mass function is principally nonperturbative on $Q^2\in[0,M_N^2]$.  

Together these observations suggest that, in our model and in QCD, a judicious estimate of the least-upper-bound on the domain of soft momenta is $\Lambda = M_N$, and this is the value employed in Fig.\,\ref{plotF2F1log}: the figure does not provide compelling evidence for Eq.\,(\ref{scaling}) on the domain for which information is currently available.  One can attempt to \textit{fit} Eq.\,(\ref{scaling}) to the calculated results or data, and \cite{belitsky} $\Lambda=\Lambda_{\rm fit}\approx 0.3\,$GeV, provides fair agreement.  However, this value is significantly smaller than the natural scales we have identified in the model and $R^0_{u}(\Lambda_{\rm fit}^2)=0.98\,R^0_{u}(0)$.  %Hence one may regard this agreement as nugatory.

\section{Epilogue}
\label{epilogue}
We recapitulated on a Poincar\'e covariant Faddeev equation that describes baryons as composites of confined-quarks and -diquarks, and solved this equation to obtain masses and amplitudes for the nucleon and $\Delta$.  Two parameters appear in the model Faddeev equation: masses of the scalar and axial-vector diquark correlations.  They were fixed by fitting stipulated masses of the baryons.  We interpreted the masses and Faddeev amplitudes thus obtained as representing properties of the baryons' ``quark core,'' and argued that this should be augmented in a consistent fashion by chiral-loop corrections.

We explained subsequently the formulation of a nucleon-photon vertex, which automatically ensures the vector Ward-Takahashi identity is fulfilled for on-shell nucleons described by the calculated Faddeev amplitudes.  This guarantees current conservation.  The vertex \textit{Ansatz} involves three parameters.  Two of these specify electromagnetic properties of axial-vector diquarks and a third measures the strength of electromagnetically induced axial-vector- $\leftrightarrow$ scalar-diquark transitions.  These quantities are also properties of the nucleons' quark core.

The elements just described are sufficient for a calculation of the quark contribution to the nucleons' electromagnetic form factors.  We explored a reasonable range of nucleon-photon-vertex parameter values and found that an accurate description of the nucleons' static properties was not possible with the core components alone.  However, this mismatch with experiment was greatly reduced by the inclusion of chiral corrections.  %Since true pseudoscalar mesons are not pointlike their contribution to baryon form factors diminishes with increasing momentum transfer.  Thus experiments on nucleons involving $Q^2 \gtrsim 2\,$GeV$^2$ probe properties of the Poincar\'e covariant quark core.  This was explicated in our study.  

We calculated ratios of the proton's form factors.  On the whole domain of nucleon-photon-vertex parameter values explored, the calculated behaviour of $G_E^p(Q^2)/G_M^p(Q^2)$ for $Q^2 \gtrsim 2\,$GeV$^2$ agrees with that inferred from contemporary polarisation transfer data.  Moreover, with the same insensitivity to parameters, the ratio $\sqrt{Q^2} F_2(Q^2)/F_1(Q^2) \approx\,$constant on $Q^2\in [2,6]\,$GeV$^2$.  Since the parameters in the nucleon-photon vertex do not influence these outcomes, we judge they are manifestations of features intrinsic to the nucleon's Faddeev amplitude.  In the nucleon's rest frame, this amplitude corresponds to a relativistic wave function with $s$-, $p$- and even $d$-wave quark orbital angular momentum components.  %On the domain of $Q^2$ we explored there is no compelling evidence for the view that $F_2(Q^2)/F_1(Q^2) \propto [\ln Q^2]^2/Q^2$.  

In our view baryons can realistically be seen as a dominant Poincar\'e covariant quark core, augmented by pseudoscalar meson cloud contributions that, e.g., make a noticeable contribution to form factors for $Q^2\lesssim 2\,$GeV$^2$.  Meson compositeness ensures that such contributions diminish with increasing $Q^2$.  Hence experiments at larger $Q^2$ serve as an instructive probe of correlations in baryon wave functions; i.e., their Faddeev amplitudes.  A good understanding of QCD's long-range dynamics is required in order to obtain a reliable quark-core wave function.  This is the box that contemporary experiments are opening.

%\hspace*{-\parindent}\begin{minipage}[t]{\textwidth}
\begin{acknowledge}
We thank J.~Arrington, P.~Maris, M.~Oettel, P.\,C.~Tandy and S.\,V.~Wright for constructive conversations.
This work was supported by: %
the Austrian Research Foundation \textit{FWF, Erwin-Schr\"odinger-Stipendium} no.\ J2233-N08; COSY contract no.\ 41445395; Department of Energy, Office of Nuclear Physics, contract no.\ W-31-109-ENG-38; \textit{Deutsche Forschungsgemeinschaft} contract no.\ GRK683 (European Graduate School T{\"u}bingen-Basel); \textit{Helmholtz-Gemeinschaft} (Virtual Theory Institute VH-VI-041); National Science Foundation contract no.\ INT-0129236; and the \textit{A.\,v.\ Humboldt-Stiftung} via a \textit{F.\,W.\ Bessel Forschungspreis}.
%; and benefited from the facilities of ANL's Computing Resource Center.
\end{acknowledge}
%\end{minipage}
%\medskip

\appendix
\section{Euclidean Conventions} 
\setcounter{section}{1}
\label{App:EM} 
In our Euclidean formulation: 
\begin{equation} 
p\cdot q=\sum_{i=1}^4 p_i q_i\,; 
\end{equation} 
% \begin{equation} 
\begin{equation}
\{\gamma_\mu,\gamma_\nu\}=2\,\delta_{\mu\nu}\,;\; 
\gamma_\mu^\dagger = \gamma_\mu\,;\; 
\sigma_{\mu\nu}= \sfrac{i}{2}[\gamma_\mu,\gamma_\nu]\,; \;
% \end{equation} 
% and 
% \begin{equation} 
{\rm tr}_[\gamma_5\gamma_\mu\gamma_\nu\gamma_\rho\gamma_\sigma]= 
-4\,\epsilon_{\mu\nu\rho\sigma}\,, \epsilon_{1234}= 1\,.  
% \end{equation} 
\end{equation}

A positive energy spinor satisfies 
\begin{equation} 
\bar u(P,s)\, (i \gamma\cdot P + M) = 0 = (i\gamma\cdot P + M)\, u(P,s)\,, 
\end{equation} 
where $s=\pm$ is the spin label.  It is normalised: 
\begin{equation} 
\bar u(P,s) \, u(P,s) = 2 M 
\end{equation} 
and may be expressed explicitly: 
\begin{equation} 
u(P,s) = \sqrt{M- i {\cal E}}\left( 
\begin{array}{l} 
\chi_s\\ 
\displaystyle \frac{\vec{\sigma}\cdot \vec{P}}{M - i {\cal E}} \chi_s 
\end{array} 
\right)\,, 
\end{equation} 
with ${\cal E} = i \sqrt{\vec{P}^2 + M^2}$, 
\begin{equation} 
\chi_+ = \left( \begin{array}{c} 1 \\ 0  \end{array}\right)\,,\; 
\chi_- = \left( \begin{array}{c} 0\\ 1  \end{array}\right)\,. 
\end{equation} 
For the free-particle spinor, $\bar u(P,s)= u(P,s)^\dagger \gamma_4$. 
 
The spinor can be used to construct a positive energy projection operator: 
\begin{equation} 
\label{Lplus} \Lambda_+(P):= \frac{1}{2 M}\,\sum_{s=\pm} \, u(P,s) \, \bar 
u(P,s) = \frac{1}{2M} \left( -i \gamma\cdot P + M\right). 
\end{equation} 
 
A negative energy spinor satisfies 
\begin{equation} 
\bar v(P,s)\,(i\gamma\cdot P - M) = 0 = (i\gamma\cdot P - M) \, v(P,s)\,, 
\end{equation} 
and possesses properties and satisfies constraints obtained via obvious analogy 
with $u(P,s)$. 
 
A charge-conjugated Bethe-Salpeter amplitude is obtained via 
\begin{equation} 
\label{chargec}
\bar\Gamma(k;P) = C^\dagger \, \Gamma(-k;P)^{\rm T}\,C\,, 
\end{equation} 
where ``T'' denotes a transposing of all matrix indices and 
$C=\gamma_2\gamma_4$ is the charge conjugation matrix, $C^\dagger=-C$. 
 
In describing the $\Delta$ resonance we employ a Rarita-Schwinger spinor to 
unambiguously represent a covariant spin-$3/2$ field.  The positive energy 
spinor is defined by the following equations: 
\begin{equation} 
\label{rarita}
(i \gamma\cdot P + M)\, u_\mu(P;r) = 0\,,\;
\gamma_\mu u_\mu(P;r) = 0\,,\;
P_\mu u_\mu(P;r) = 0\,, 
\end{equation} 
where $r=-3/2,-1/2,1/2,3/2$.  It is normalised: 
\begin{equation} 
\bar u_{\mu}(P;r^\prime) \, u_\mu(P;r) = 2 M\,, 
\end{equation} 
and satisfies a completeness relation 
\begin{equation} 
\frac{1}{2 M}\sum_{r=-3/2}^{3/2} u_\mu(P;r)\,\bar u_\nu(P;r) = 
\Lambda_+(P)\,R_{\mu\nu}\,, 
\end{equation} 
where 
\begin{equation} 
R_{\mu\nu} = \delta_{\mu\nu} I_{\rm D} -\frac{1}{3} \gamma_\mu \gamma_\nu + 
\frac{2}{3} \hat P_\mu \hat P_\nu I_{\rm D} - i\frac{1}{3} [ \hat P_\mu 
\gamma_\nu - \hat P_\nu \gamma_\mu]\,, 
\end{equation} 
with $\hat P^2 = -1$, which is very useful in simplifying the positive energy 
$\Delta$'s Faddeev equation. 

\section{Nucleon-Photon Vertex} 
\label{appB}
In order to explicate this vertex, we write the scalar and axial-vector components of the nucleons' Faddeev amplitudes in the form [cf.\ Eq.\,(\ref{FEone})]
\begin{equation}
\label{NucWF}
\Psi(k;P) = \left[
\begin{array}{l}
\Psi^0(k;P)\\
\Psi_{\mu}^{i}(k;P)
\end{array}
\right]
= \left[ 
\begin{array}{l}
\mathcal{S}(k;P) u(P)\\
\mathcal{A}_{\mu}^{i}(k;P)u(P)
\end{array}
\right],
\qquad i=1,\ldots,4\,.
\end{equation}
For explicit calculations, we work in the Breit frame: $P_\mu=P_\mu^{BF}-Q_\mu /2$, $P'_\mu=P_\mu^{BF}+Q_\mu /2$ and $P_\mu^{BF}=(0,0,0,i\sqrt{M_n^2+Q^2/4})$, and write the electromagnetic current matrix element as [cf.\ Eq.\,(\ref{Jnucleon})]
\begin{eqnarray}
\label{ABcurrent}
\left\langle P' | \hat{J}_{\mu}^{em} | P \right\rangle
&=& \Lambda^+(P')
\left[ \gamma_\mu G_E + M_n \frac{P_{\mu}^{BF}}{P_{BF}^{2}}
(G_E-G_M) \right] \Lambda^+(P)
\\
&=& \int \frac{d^4 p}{(2\pi)^4}\,\frac{d^4 k}{(2\pi)^4}\,
\bar{\Psi}(-p,P') J_{\mu}^{em}(p,P';k,P) \Psi(k,P)\,.
\end{eqnarray}
In Fig.\,\ref{vertex} we have broken the current, $J_{\mu}^{em}(p,P';k,P)$,  into a sum of six parts, each of which we now make precise.

The one-body term of Diagram~1 is expressed as 
\begin{eqnarray}
\label{B1}
J_{\mu}^{qu} &=&  S(p_q) \hat{\Gamma}_{\mu}^{qu}(p_q;k_q) S(k_q) 
\left(\Delta^{0^+}(k_s) + \Delta^{1^+}(k_s) \right)
(2\pi)^4 \delta^4(p-k-\hat{\eta}Q)\,,
\end{eqnarray}
where $\hat\Gamma_{\mu}^{qu}(p_q;k_q)= Q_q \, \Gamma_{\mu}(p_q;k_q)$, with $Q_q={\rm diag}[2/3,-1/3]$ being the quark electric charge matrix, and $\Gamma_{\mu}(p_q;k_q)$ is given in Eq.\,(\ref{bcvtx}).

Diagrams~2 and 4 are obtained through 
\begin{eqnarray}
\label{B2}
J_{\mu}^{dq} &=& \Delta^i(p_{d}) 
\left[ \hat{\Gamma}_{\mu}^{dq}(p_{d};k_{d}) \right]^{i j} 
\Delta^{j}(k_{d}) S(k_q)
(2\pi)^4 \delta^4(p-k+\eta Q)\,.
\end{eqnarray}
For Diagram~2: $ [\hat{\Gamma}_{\mu}^{dq}(p_{d};k_{d})]^{i j}={\rm diag}[Q_{0^+} \Gamma_\mu^{0^+},Q_{1^+}\Gamma_\mu^{1^+}] $, where $Q_{0^+}=1/3$ and $\Gamma_\mu^{0^+}$ is given in Eq.\,(\ref{Gamma0plus}), and $Q_{1^+}={\rm diag}[4/3,1/3,-2/3]$ with $\Gamma_\mu^{1^+}$ given in Eq.\,(\ref{AXDQGam}); while for Diagram~4: $[\hat{\Gamma}_{\mu}^{dq}(p_{d};k_{d})]^{i= j}=0$, and $[\hat{\Gamma}_{\mu}^{dq}(p_{d};k_{d})]^{1,2}=\Gamma_{SA}$, which is given in Eq.\,(\ref{SAPhotVertex}), and $[\hat{\Gamma}_{\mu}^{dq}(p_{d};k_{d})]^{2,1}=\Gamma_{AS}$.  Naturally, the diquark propagators match the line to which they are attached. 

In Eqs.\,(\ref{B1}) \& (\ref{B2}) the momenta are
\begin{eqnarray}
\label{etavalue}
\begin{array}{lc@{\qquad}l}
k_q=\eta P+k\,, & & p_q=\eta P'+p\,, \\
k_d=\hat{\eta}P-k\,, & & p_d=\hat{\eta}P'-p\,,
\end{array}
\end{eqnarray}
with $\eta + \hat{\eta}=1$.  The results reported herein were obtained with $\eta=1/3$, which provides a single quark with one-third of the baryon's total momentum, but, as our approach is manifestly Poincar\'e covariant, the precise value is immaterial.  Nevertheless, numerical results converge more quickly with this natural choice. 

The quark exchange contribution to the vertex, Diagram~3, is obtained with
\begin{equation}
\label{B3}
J_{\mu}^{ex} = -\frac{1}{2} S(k_{q}) \Delta^{i}(k_{d})
\Gamma^{i}(p_1,k_{d})
S^T(q) \hat{\Gamma}_{\mu}^{quT}(q',q) S^T(q')
\bar{\Gamma}^{jT}(p'_2,p_{d})
\Delta^j(p_{d}) S(p_{q})\,,
\end{equation}
wherein the vertex $\hat{\Gamma}_{\mu}^{qu}$ appeared in Eq.\,(\ref{B1}).  The full contribution is obtained by summing over the superscripts $i,j$, which can each take the values $0^+$, $1^+$.  The so-called seagull contributions, Diagrams~5 \& 6, are given by 
\begin{eqnarray}
\label{B5}
J_{\mu}^{sg} &=& \frac{1}{2} S(k_{q}) \Delta^{i}(k_{d}) 
\left( X_{\mu}^{i}(p_q,q',k_d) S^T(q')
\bar{\Gamma}^{jT}(p'_2,p_{d})
\right.
\nonumber\\
& & -
\left. 
\Gamma^{i}(p_1,k_{d}) S^T(q) 
\bar{X}_{\mu}^{j}(-k_q,-q,p_d)
\right) \Delta^{j}(p_{d}) S(p_{q})\,,
\end{eqnarray}
where, again, the superscripts are summed.  In Eqs.\,(\ref{B3}) \& (\ref{B5}) the momenta are 
\begin{eqnarray}
\begin{array}{lc@{\qquad}l}
q = \hat{\eta}P-\eta P'-p-k\,, & & q' = \hat{\eta}P'-\eta P-p-k \,,\\
p_1 = (p_q-q)/2\,,& & p'_2 = (-k_q+q')/2 \,,\\
p'_1 = (p_q-q')/2\,, & & p_2 = (-k_q+q)/2 \,.
\end{array}
\end{eqnarray}

Diagrams~1, 2 and 4 are one-loop integrals, which we evaluated by Gau{\ss}ian quadrature.  The remainder, Diagrams 3, 5 and 6, are two-loop integrals, for whose evaluation we employed Monte-Carlo methods.

%------------------------------------------------------------------------ 

\end{document}